\documentclass[a4paper,10pt]{article}
\usepackage[pdftex]{graphicx,hyperref}
\usepackage{amsmath,amsfonts,amssymb}
\usepackage{wasysym}
\usepackage{fullpage}
\usepackage{enumerate}
\usepackage{sidecap}
\usepackage{lscape}
\usepackage{authblk}
\usepackage{appendix}
\usepackage{setspace}
\usepackage{enumitem}
\usepackage[table, x11names]{xcolor}

\usepackage[utf8]{inputenc}

\usepackage[T1]{fontenc}
\usepackage{amsthm}
\usepackage{amsmath}
\usepackage{bbm}

\theoremstyle{definition}
\newtheorem{definition}{Definition}
\newtheorem{lemma}{Lemma}
\newtheorem{corollary}{Corollary}
\newtheorem{theorem}{Theorem}
\newtheorem{remark}{Remark}

\newtheorem{claim}{Claim}

\newtheorem*{notation}{Notation}

\newcommand{\E}{\mathbb E}
\renewcommand{\P}{\mathbb P}

\newcommand{\ptaumOn}{\frac{1}{\tau-1}}
\newcommand{\ptaumTw}{\frac{1}{\tau-2}}
\newcommand{\ptaumTh}{\frac{1}{|\tau-3|}}
\newcommand{\pThmtau}{\frac{1}{3-\tau}}

\newcommand{\yellow}{\cellcolor[HTML]{FFFDCA}}
\newcommand{\green}{\cellcolor[HTML]{D6FFD5}}
\newcommand{\blue}{\cellcolor[HTML]{E6FFF5}}
\newcommand{\orange}{\cellcolor[HTML]{FEE8CD}}
\newcommand{\red}{\cellcolor[HTML]{FFD1C6}}
\newcommand{\lgrey}{\cellcolor[HTML]{F5F5F5}}
\newcommand{\dgrey}{\cellcolor[HTML]{D3D3D3}}

\title{\textbf{Switchover phenomenon induced by epidemic seeding on geometric networks}}

\date{}

\author[a]{Gergely \'Odor}
\author[b]{Domonkos Czifra} 
\author[c]{J\'ulia Komj\'athy}
\author[b,d,*]{\\ L\'aszl\'o Lov\'asz}
\author[e,b]{M\'arton Karsai}

\affil[a]{\small \'Ecole polytechnique f\'ed\'erale de Lausanne, CH-1015 Lausanne, Switzerland}
\affil[b]{Alfr\'ed R\'enyi Insititute of Mathematics, H-1053 Budapest, Hungary}
\affil[c]{Eindhoven University of Technology, NL-5612AZ Eindhoven, The Netherlands}
\affil[d]{E\"otv\"os Lor\'ant University, H-1053 Budapest, Hungary}
\affil[e]{Department of Network and Data Science, Central European University, A-1100 Vienna, Austria}
\affil[*]{Correspondence should be addressed to L\'aszl\'o Lov\'asz (laszlo.lovasz@ttk.elte.hu)}

\begin{document}

\maketitle

\begin{abstract}
It is a fundamental question in disease modelling how the initial seeding of an epidemic, spreading over a network, determines its final outcome. Research in this topic has primarily concentrated on finding the seed configuration which infects the most individuals. Although these optimal configurations give insight into how the initial state affects the outcome of an epidemic, they are unlikely to occur in real life. In this paper we identify two important seeding scenarios, both motivated by historical data, that reveal a new complex phenomenon. In one scenario, the seeds are concentrated on the central nodes of a network, while in the second, they are spread uniformly in the population. Comparing the final size of the epidemic started from these two initial conditions through data-driven and synthetic simulations on real and modelled geometric metapopulation networks, we find evidence for a switchover phenomenon: When the basic reproduction number $R_0$ is close to its critical value, more individuals become infected in the first seeding scenario, but for larger values of $R_0$, the second scenario is more dangerous. We find that the switchover phenomenon is amplified by the geometric nature of the underlying network, and confirm our results via mathematically rigorous proofs, by mapping the network epidemic processes to bond percolation. Our results expand on the previous finding that in case of a single seed, the first scenario is always more dangerous, and further our understanding why the sizes of consecutive waves can differ even if their epidemic characters are similar.
\end{abstract}

\section{Introduction}

Whether a local epidemic becomes a global pandemic depends on several conditions. Biological~\cite{pastor2015epidemic}, environmental~\cite{stone2007seasonal} and behavioral~\cite{funk2009spread} factors are important but the final outcome of the epidemic is also strongly determined by the size and location of the seed population where it originates from~\cite{memish2021tale,colizza2007invasion,apolloni2014metapopulation,Kitsak_2018_influential_spreaders}. If the epidemic strikes first at an isolated place with low population density and few local transportation connections, it may become rapidly extinct without causing a major breakout. The dynamics can be entirely different if the epidemic starts from a well connected, more populated place where it can survive and spread to the rest of the population more easily. Although this is the broadly accepted picture, we challenge this intuition and show that seeding an epidemic from the most tightly connected core of a network does not always lead to a larger epidemic in the long run: If the disease transmits easily, seeding the spreading from nodes selected uniformly at random from the network could reach a larger population. 

\begin{figure*}[t!]
    \centering
    \includegraphics[width = 1.\textwidth]{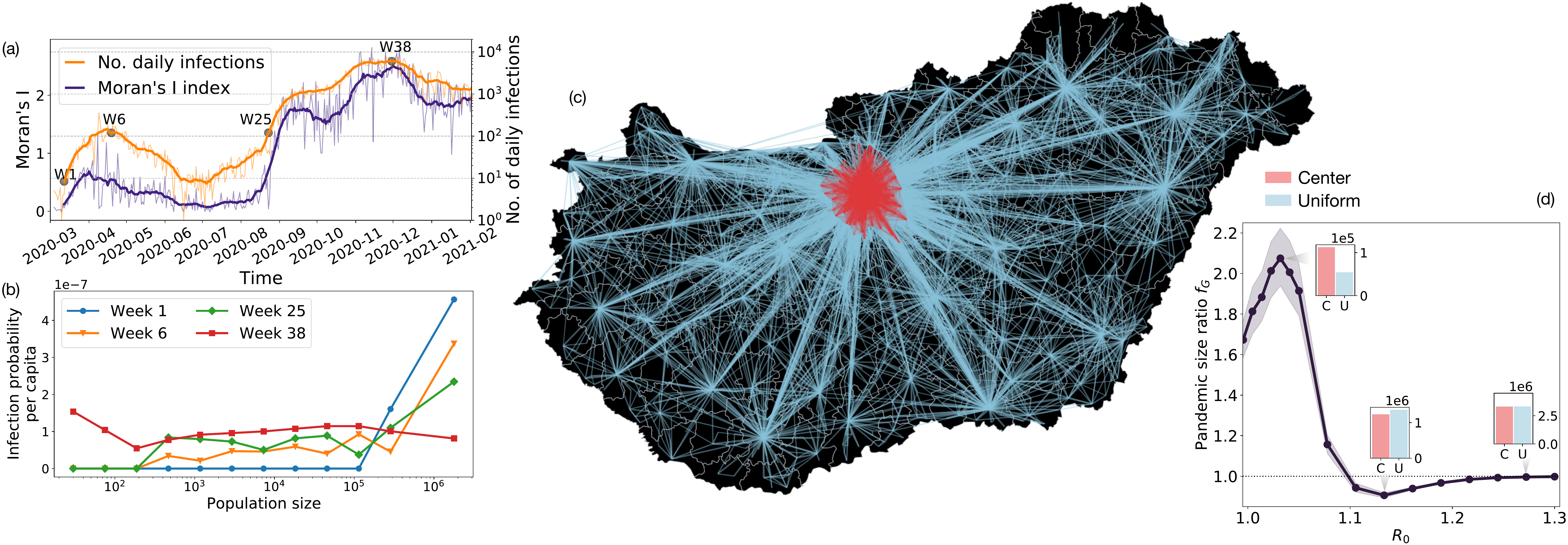}
    \caption{\small Data-driven observations of the switchover phenomenon. (a) Dynamics of the number of daily infections (orange) and the Moran's I index (purple) for Hungary. Indicated time points match the observation weeks in panel b. (b) Distribution of per capita infection probabilities in settlements of different sizes at different observation times (in weeks). (c) Commuting network map of Hungary with settlements larger than $1000$ inhabitants and commuting links with more than 25 travelers depicted. Central Hungary (called Center) is highlighted with red. (d) Pandemic size ratios $f_G(R_0, s)$ measured between the endemic sizes of simulated SIR epidemic processes seeded from $s$ populations selected from the center or uniformly at random from the whole metapopulation network. Epidemic seeded from the center may lead to larger outbreaks for small $R_0$ basic reproduction numbers (left bar plot), while uniform seeding results more infections for larger $R_0$ (middle bar plot). For very large $R_0$ values, differences due to different seeding strategies disappear (right bar plot).}
    \label{fig:1}
\end{figure*}

Similar phenomena could act in the background during the early phase of the COVID-19 pandemic: Even though the circulating SARS Cov-2 epidemic variants had similar transmission profiles, the number of infections differed significantly in subsequent waves of the pandemic in several countries~\cite{brauner2021inferring,dye2020scale,white2020state}. This was especially true for Hungary with an order of magnitude more daily number of detected cases observed at the peak of the second wave as compared to the first outbreak (see Figure~\ref{fig:1}a). Reasons behind this variation could be the effect of several factors. This includes seasonal effects as people may have spent more time outside during the first wave~\cite{merow2020seasonality}; Regulations were followed less strictly during the second wave that may have potentially induced a larger number of contacts per person transmitting the disease~\cite{reicher2021pandemic}; The testing capacities also developed considerably since the beginning of the pandemic, allowing for more observations during the second wave; Further, while the first-wave of the epidemic was boosted by institutional outbreaks (e.g.\ in hospitals and care homes) that were easier to identify and contain~\cite{burton2020evolution}, the second wave circulated freely in the population without effective control~\cite{aleta2020modelling}.

The global and local mobility of people are among the most important driving factors behind the spatial spread of most diseases~\cite{kraemer2020effect,bajardi2011human}. How people travel locally as well as between cities and countries can be well represented as geometric metapopulation networks~\cite{colizza2007reaction}, where nodes are populations (cities) and (weighted) links code the number of people traveling between them for different purposes. Concentrating on Hungary, we consider a spatial mobility network (see Fig.~\ref{fig:1}c) describing the average number of daily commuters to work and school between $1398$ settlements with populations larger than $1000$ according to the 2016 Hungarian microcensus~\cite{hungary2016census}. From epidemic data we can follow the daily number of new COVID-19 infection cases in each of these settlements to explore their spatio-temporal distribution in this geometric network. The analysis of the epidemic on this structure sheds light on a so-far neglected effect associated to the different initial seeding conditions of the virus, which may contributed to the emerging large differences between the first and the second waves.

The first wave started in March 2020 in Hungary (W1 in Fig.~\ref{fig:1}a). As in many countries, the disease arrived to the country via international air-travel and first landed in larger cities~\cite{karsai2020hungary,fauver2020coast,kang2020spatial} resulting in outbreaks clumped around highly populated areas. This is evident from Fig.~\ref{fig:1}b, where the per-capita infection probability at the beginning of the first wave (week 1) indicates that infection cases were concentrated in cities with the largest populations. To further demonstrate how much of the infection spreading can be attributed to everyday mobility (as opposed to atypical mobility patterns, such as going on a vacation), we computed the Moran's I index on this network (for definition see Methods). This is a spatial auto-correlation function, which has been previously used to measure the spatial association of the COVID-19 infections by~\cite{kang2020spatial}. Looking at the time dependency of the Moran's I index (on Figure~\ref{fig:1}a), during the beginning of the first wave (W1) the index indicates low spatial correlation, meaning that infected cases were concentrated only in a few places during this initial stage of the epidemic. In contrast, the second wave in Hungary (and Europe) emerged after the summer season, and was potentially induced by people coming back from holidays bringing back the virus to their local community, and thus re-starting the pandemic from a significantly different initial condition. Indeed, at the beginning of the second wave (at the end of August 2020 in Hungary, see Fig.~\ref{fig:1}a) new infected cases were distributed more homogeneously all around the country. On the one hand, this is evident from Fig.~\ref{fig:1}b where the corresponding probability distribution (week 25) is more stretched towards smaller population, as compared to week 1. On the other hand, the same conclusion can be drawn from Fig.~\ref{fig:1}a (W25) where the Moran's I index starts to grow rapidly from a state where infections were even more homogeneously distributed than at the peak of the first wave (W6), although the infection numbers were comparable. This homogenization of infected cases continued during the unfolding of the second wave leading to a fully homogeneous distribution -- corresponding to population densities -- at the peak (W38 in Fig.~\ref{fig:1}a). Surprisingly, the first wave that started from the most tightly connected and largest populations lead to significantly smaller number of infections as compared to the second wave, that reached an order of magnitude more people, even though it was initiated from more uniformly distributed populatioms of the network.

To capture better this structural distinction of the spatial commuting network of Hungary, 
we identify a \emph{central node set} ($\mathcal C$) containing Budapest and its suburbs (with about 30\% of the population of the country)~\cite{centralhungary} (indicated with red in Fig.~\ref{fig:1}c), as a subset of all settlements $V$ in the country. To capture the two seeding conditions that we observed earlier, we simulated Susceptible-Infected Recovered (SIR) model processes on the metapopulation network (for definitions see next section and Methods) with given $\beta$ infection and $\mu$ recovery rates determining the basic reproduction number $R_0=\beta / \mu$ of the process (the average number of people infected by one ill person in a susceptible population). We considered two initial conditions by initiating the SIR process from the same number of seeds distributed among the populations in $\mathcal C$ or uniformly at random in populations in $V$ from the whole country. To observe the relative effects of these seedings, we look at the \emph{experimental pandemic size ratio} $f_G(R_0, s)$ of the final infection size of processes seeded from $s$ populations from the central set divided by that when seeding randomly from the whole country (for a related but more formal definition see section on Theoretical results below). Interestingly, as shown in Fig.~\ref{fig:1}d, while $R_0\simeq 1$ is small and close to the critical point for a global outbreak ($R_0>R_c^{\mathrm{glob}}>1)$, we find $f_G(R_0, s)>1$, thus the epidemic seeded from the central population set leads to larger outbreaks. However, as we increase $R_0$, the fraction $f_G(R_0, s)$ falls under $1$, thus seeding from uniformly random selected populations over the whole country induces a larger outbreak. Finally as $R_0$ grows even larger, the difference between these seeding scenarios vanishes as the epidemic reaches essentially the whole population in each case.
In this paper we study this switchover phenomenon between the two seeding scenarios in focus and argue that the geometric nature of the underlying network plays an important role in amplifying these effects. We perform data-driven and synthetic simulations of spreading processes on real, geometric, and random metapopulation networks and provide a rigorous proof of the phenomenon after mapping it to a bond percolation problem. These observations challenge the commonly accepted intuition suggesting that the size of the epidemic is always the largest if seeded from the best connected sub-graph or from the largest degree nodes of a network. We give rigorous proofs that this phenomenon appears in various random networks.

\begin{figure*}[t!]
    \centering
    \includegraphics[width = 1.\textwidth]{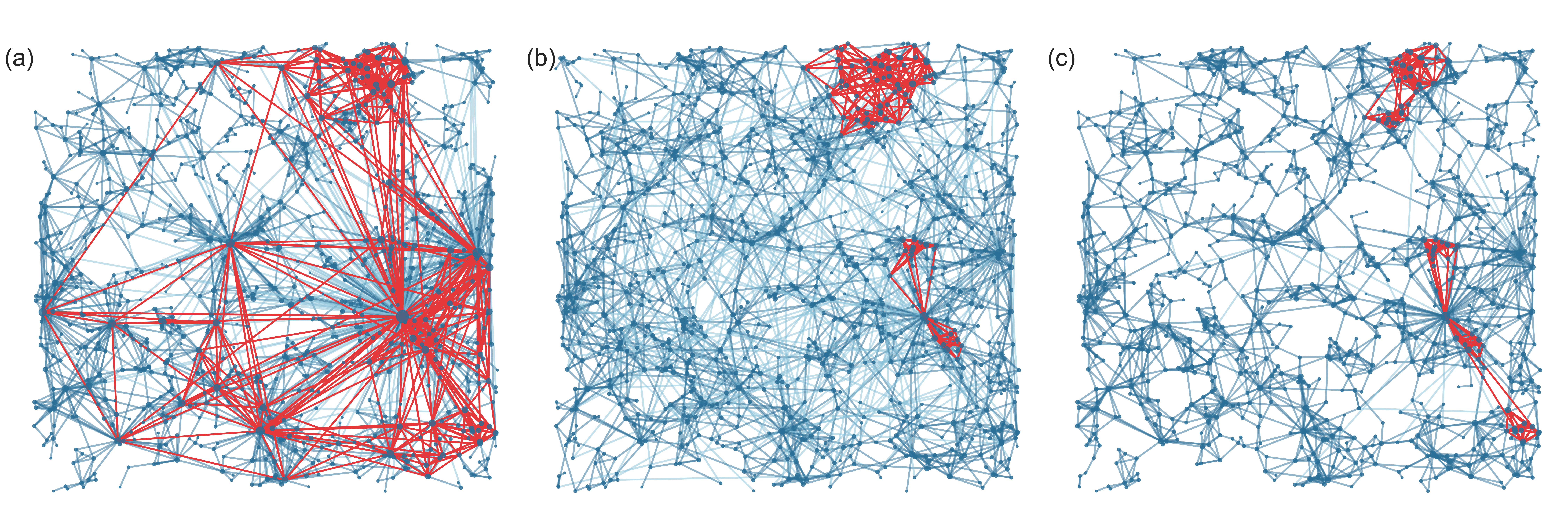}
    \caption{\small Geometric Inhomogenous Random Graph (GIRG) models. (a) Model networks with connection parameter $\tau=2.5$ and geometry parameter $\alpha=2.3$. In case $\tau<3$ the network appears with high degree variability and dominant hubs (light blue) connected via long-range edges. (b) By increasing $\tau>3$ (here $\tau=3.5$) and decreasing $\alpha=1.3$, hubs' sizes reduce and long-range interactions become more random. (c) With parameter $\alpha=2.3$ (and $\tau=3.5$) the network is strongly geometric with (dark blue) short-range interactions and no long-range links. Networks were generated over the same $N=1000$ nodes randomly distributed in a unit square and the highest k-cores of each graph are colored in red. Red clusters highlight the largest k-cores in each network.}
    \label{fig:2}
\end{figure*}

\section*{Results}

Metapopulation network models of epidemic spreading~\cite{colizza2007reaction} have been proven to be useful for precise simulations of real-world epidemic phenomena, while in some simple cases they are treatable even analytically. They are defined on a network $G=(V,E)$ where nodes $v\in V$ are populations of size $n_v$ and links $e(u,v,w)\in E$ code the $w$ number of movements between connected populations $u$ and $v$. An epidemic spreading, like an SIR model, on this network can be interpreted as a \emph{reaction-diffusion process}~\cite{colizza2007reaction} where individuals can be in one of three mutually exclusive states ($S$-susceptible, $I$-infected or $R$-recovered). In one iteration, during the reaction phase, individuals in the same population mix homogeneously and possibly pass the infection with rate $\beta$ between infected and susceptible ones, or if actually infected, they may recover with rate $\mu$ reaching an absorbing state ($R$), disabling them to get infected again. Subsequently, during the diffusion phase, individuals (possibly infected) may move to neighboring nodes in the metapopulation network, this way migrating the epidemic to other populations (for a more formal definition see Methods). To capture commuting behavior in our system, we assume that every individual has a `home' population. At each iteration step, each individual $i$ in each population is being selected for moving with probability $p_m$. If selected, $i$ migrates to a neighboring population selected by a probability proportional to the link weights, while $i$ returns to its home in the subsequent iteration step.

\subsection*{Simulation results}\label{sec:sim_results}
The nodes and links in a metapopulation network model can be respectively associated with real settlements and commuting links in a country. This defines a spatially embedded geometric network (see Fig.~\ref{fig:1}c for Hungary) featuring various structural heterogeneities (for a detailed data description see Methods). Geometric constraints inducing commuting connections at various distances, link weights coding the daily commuting frequencies between populations, the number of commuting connections of each settlement (also called the node degree in the network), or the size of the different populations are all network characteristics taking values ranging over orders of magnitudes. These properties may all contribute to the emergence of the observed switchover phenomenon of simulated spreading processes (an SIR model in our case), with central vs random seeding in the meta-network.

\begin{figure*}[t!]
    \centering
    \includegraphics[width = 1.\textwidth]{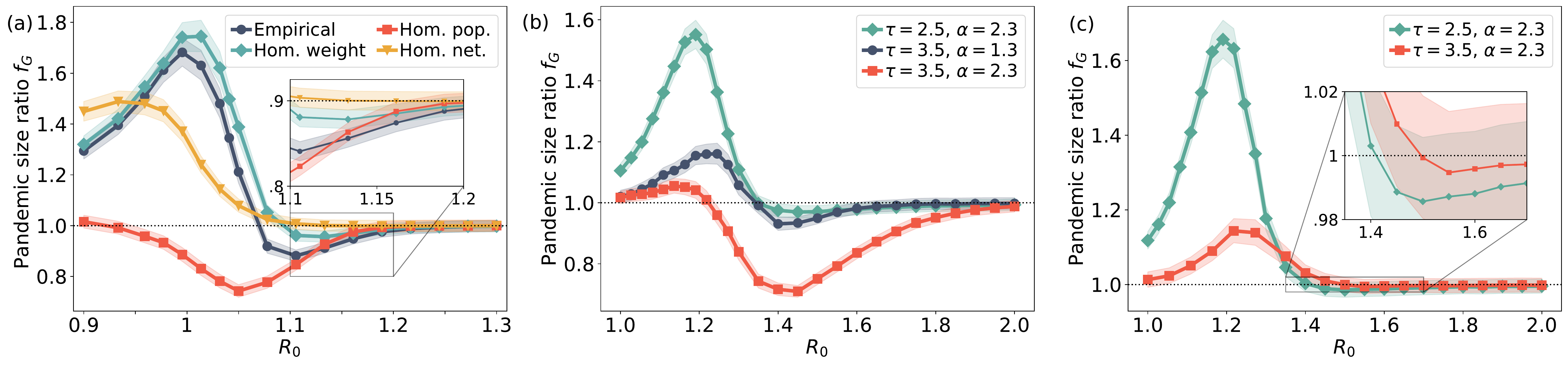}
    \caption{\small The pandemic size ratio $f_G$ as a function of $R_0$.
    a) Simulation results on real commuting network of Hungary and its three homogenized versions as explained in the main text. Each data point is an average computed from 150 independent simulations, shown with $81\%$ confidence interval. For each of them initially 97 settlements are infected, distributing $0.0005\%$ infected agents uniformly at random in the simulated population of $10^7$.
    b) Three geometric inhomogeneous random graph models corresponding to the three main universality classes with respect to graph distance. When $(\tau, \alpha)=(2.5, 2.3)$, the core is de-localized, the switchover is weak. When $(\tau, \alpha)=(3.5, 1.3)$ or $(3.5, 2.3)$, the underlying geometry is more apparent, the core is localized, so the parameter range for $R_0$ where random seeding is more dangerous ($f_G<1$) is more spread-out. On c), we see $f_G$ on the configuration model, where weak switchover emerges. For b),c) the size of the metapopulation networks are $n=1000$ and each population is set with N=2000 individuals. Each pandemic size ration data point is computed on 25 networks, with 35 simulations, distributing $0.0005\%$ infected agents uniformly at random between 30 settlements.}
    \label{fig:3}
\end{figure*}

To identify which underlying network characteristics are the most important to induce the switchover phenomenon, we use random reference network models~\cite{newman2018networks}. We homogenize the network in different ways to remove certain structural heterogeneities, and compare the outcome of simulated spreading processes on the randomized structures to our observations on the empirical network (see blue dotted curve in Fig~\ref{fig:3}a). First, to reduce the effects of weight hegerogeneities, we reset edge weights to the mean weight of all outgoing edges of each population (see green diamond curve in Fig~\ref{fig:3}a). Although this way of homogenization changes somewhat the pandemic size ratio function, it does not have dramatic effects on the observed phenomena. Second, to remove the effects of heterogeneous population sizes and the varying fraction of commuting individuals from different settlements, we set each population to the system average (6581) and choose the fraction of commuters to be the same (0.184) for each population. Interestingly, this way of homogenization makes the switchover phenomenon even stronger (see red squared curve in Fig~\ref{fig:3}a). Finally, we re-shuffle the ends of network links using the configuration network model~\cite{newman2018networks}. This  removes any structural correlations from the network beyond degree heterogeneity, including geometric effects such as long distance connections, the central-periphery structure, structural hierarchy, and locally dense sub-graphs. Due to this shuffling process the switchover phenomenon disappears (see yellow triangle curve in Fig~\ref{fig:3}a), indicating that geometric correlations play a central role behind its emergence.

\paragraph{Geometric Inhomogenous Random Graphs.} The specific effect of an underlying geometry can be studied by using geometric network models, opening directions for an analytical description of the phenomenon. Geometric Inhomogenous Random Graph (GIRG) models~\cite{Sampling_GIRG} provide a good framework to generate structurally heterogeneous synthetic metapopulation structures embedded in geometric space (for detailed definition see Methods). GIRGs have two robust parameters that control the qualitative features of the emerging network. The parameter~$\tau$ determines the variability of the number of neighbors of individual nodes (smaller values of $\tau$ correspond to more variability, while keeping the average degree the same). This is apparent when comparing Fig.~\ref{fig:2}a to  \ref{fig:2}c where all parameters of the simulated network structures are identical only $\tau$ is increased gradually, leading to the disappearance of hubs, i.e., nodes with large number of neighbors. The other robust parameter of GIRG, $\alpha$ controls the number of long-range connections in the network coding the possible travels between far-apart populations. If $\alpha\simeq 1$, many long-range edges appear resembling an ageometric (or mean-field) structure (see Fig.~\ref{fig:2}b), but when $\alpha$ is increased, the number of long-range contacts are reduced, and the network exhibits a more apparent underlying geometry as demonstrated in Fig.~\ref{fig:2}c. The values of $(\tau,\alpha)$ determine different \emph{universality classes} of GIRGs with respect to average distance in the network (see more detailed definition and explanation in Methods). 

To distinguish between the central set $\mathcal C$ from the rest of the network  we adopt the concept of \emph{core decomposition}. Formally, this procedure provides the highest core as a sub-graph of size at least $s$ with nodes having at least $k$ neighbors inside the core, for the largest possible $k$ (for definition see Methods). Similarly to the data-driven simulations, we start the spreading process from two seeding conditions: by initially selecting $s$ populations within the highest core of the metapopulation network (corresponding to the central set $\mathcal C$), or by selecting the same number of populations uniformly at random from the whole structure, and infecting $m$ individuals in each selected population in both scenarios.

We observe a similar but stronger switchover  phenomena of the pandemic size ratio $f_G$ in GIRGs as compared to the data-driven simulations. As seen in Fig.~\ref{fig:3}b, the ``shape'' of  $f_G(R_0,s)$ as the function of $R_0$ strongly depends on the network properties controlled by the parameters of the model. If the network parameter $\tau\ge 3$, the modeled epidemic processes, which were initiated from uniform random seeds reached larger populations, reflected by $f_G$ (blue curve in Fig.~\ref{fig:3}b) falling well below 1 for a broad range of $R_0$. This is because the hubs in the network have relatively smaller degrees compared to $\tau<3$. They are too far away from each other to form direct connections, thus the highest cores are localized around some of them as demonstrated in Fig.~\ref{fig:2}b and c. Although high degree seed nodes in these cores should have an advantage to effectively induce a larger outbreak, this effect is not strong enough to compensate for the disadvantage of starting the infection from a localized setup.
Beyond localized cores, long range interactions also have important effects on the network structure. Rare long range connections (induced by higher $\alpha$ values) reduce the number of edges leaving the localized cores, which leads to networks with dominant local geometric structures (as shown in Fig.~\ref{fig:2}c). This makes it even harder for the infection to spread from a localized setup. Thus, for $\tau\ge 3$, increasing $\alpha$ enhances the danger of the random seeding scenario, as evident from Fig.~\ref{fig:3}b where the (red) curve with $\tau=3.5$ and $\alpha=2.3$ reaches more below one than a similar curve with $\alpha=1.3$. Finally, when $\tau\in(2,3)$ (see green curve in Fig.~\ref{fig:3}b), the pandemic size ratio $f_G$ goes well above 1 for $R_0 \simeq 1$ values, and goes barely below $1$ for larger $R_0$. In this case the highest degree nodes are so dominant that they connect to each other even when they are spatially remote, this way they induce a de-localized core (see Fig.~\ref{fig:2}a). Simulations on such networks with de-localised cores resemble the phenomenon that is closest to our data-driven simulations (see Fig~\ref{fig:3}a) where the effects of the geometry are somewhat reduced due to the inter-connectedness of larger cities all over a country. For  $\tau\in (2,3)$, the parameter $\alpha$ does not have a significant effect on the network structure. 

For comparison, we also study the phenomenon on meta-networks sampled from the \emph{configuration model}, a uniform distribution over networks with a given power-law degree sequence. The configuration model has no underlying geometry and features heterogeneity only in its degree distribution, parametrized by the exponent of the power-law distribution $\tau$ (see details in Methods). For a fair comparison with the results obtained on GIRGs, we take $\tau=2.5$ to obtain a configuration model with plenty of hubs, and $\tau=3.5$ for a model with reduced degree heterogeneity. These cases correspond to two different universality classes (in both GIRGs and in configuration models) with respect to average distance (see Methods). To keep the average degree and the number of nodes  the same for the configuration model networks as in GIRGs, we obtain them by swapping randomly the links of the GIRG structures, while keeping the total number of connections for each node in tact. Interestingly, when larger hubs are present in the structure (the case of $\tau=2.5$ in Fig.~\ref{fig:3}c) the switchover phenomenon is recovered, even though the structure is fully uncorrelated. However, the switchover appears weaker, similar to the case on GIRGs where the effect of the geometry is suppressed due to the high inter-connectedness of the network. We provide a heuristic explanation of this observation during the derivation of our theoretical results below. In summary, our simulation results demonstrate that while the emergence of the switchover phenomenon requires only degree heterogeneity in the network, it is certainly amplified by geometric correlations of the underlying structure.

\subsection*{Theoretical results}
To explain rigorously the switchover phenomena we developed a mathematical framework relying on percolation theory.

\subsubsection*{Epidemics and percolation on metapopulation networks}
The pandemic size (i.e. the final number of recovered individuals) of a SIR model with deterministic, unit recovery time (e.g. a day) on a (non-meta) network $G$ has a useful connection with the commonly used simple mathematical framework of \textit{bond percolation}. In such a SIR model, every edge of the network $G$ transmits the disease at most once, when one endpoint is infected but the other is still susceptible. Equivalently, one may decide about every edge \emph{in advance}, independently with probability $p$, whether it will do so. This is called \emph{retaining the edge}, and $p$ is then the retention probability of the model. The retained edges form the percolated random subgraph $G^p$ of $G$. If a set $S$ of nodes is selected as infected seeds in the network, then the epidemic will spread exactly over the connected components (also called clusters) of $G^p$ that contain at least one node of $S$.

Metapopulation models are more difficult to treat mathematically, but a fundamental result by \cite{barthelemy2010fluctuation,colizza2008epidemic} connects the behavior of SIR on metapopulation models to bond percolation. Following their arguments, once a large outbreak occurs in a population A, the proportion of infected people within the population \emph{concentrates} around some $r_\infty\in(0,1)$ (called local outbreak ratio). Infected people during the local pandemic carry the infection to a neighboring population B and cause a large outbreak there with a certain -- computable -- probability:
\begin{equation}
    \label{eq:equivalence}
    p=1-\mathrm{exp}\left(-\frac{Np_mr_\infty\left(1-\frac{1}{R_0}\right)}{\mu} \right),
\end{equation}
where $N$ is the size of each population.
Since herd immunity is reached in each population after the first large local epidemic outbreak of size $r_\infty N$, later infections to a population are no longer able to cause macroscopically visible outbreaks. Therefore, after time-rescaling, the populations themselves go through an $S \rightarrow I \rightarrow R$ progression with unit recovery times and infection probability $p$. Consequently, the metapopulation model can be approximated by a simple SIR model on the network of populations, and in turn with a bond percolation process with retention probability $p$.

The connection between metapopulation models and bond percolation allows us to understand the switchover phenomenon of the pandemic size ratio using a theoretical analysis of percolation cluster sizes, which have been extensively studied both in the mathematics and physics literature for various network models, because they show a remarkable \emph{phase transition} in the edge retention probability $p$. At a \emph{critical value} $p_c$ two phases are separated, where for $p<p_c$ all clusters are small, while for $p>p_c$ a single \emph{giant cluster} emerges that contains a positive proportion of all nodes, while all other clusters are small. The critical parameter $p_c$ depends only on the structure of the network $G$. For some networks, $p_c$ can only be measured using numerical simulations. However, for the configuration model, the critical $p_c$ can be explicitly computed, given the degree-distribution of the network, as $p_c=\mathbb E[\deg(v)]/\mathbb E[\deg(v)(\deg (v)-1)]$, which is asymptotically nonzero when $\tau>3$. Using equation \eqref{eq:equivalence} the critical parameter $p_c$ translates back to a critical basic reproduction number $R_{c}^{\mathrm{glob}}>1$ for the infection process. For $R_0<1$ the epidemic is sub-critical already within a single population, while for $1<R_0<R_{c}^{\mathrm{glob}}$ the epidemic is super-critical within populations but sub-critical globally in the meta-network (hence outbreaks containing only a few populations are possible). Finally, for $R_0>R_{c}^{\mathrm{glob}}$ the epidemic is super-critical in the entire network.

Beyond percolation cluster sizes, we also need to understand how the different seedings (central or uniform) interact with the clusters to explain the switchover phenomenon of the pandemic size ratio.
Slightly deviating from the experimental setup, where central seeding corresponded to the highest core, here we define the central seeding set $\mathcal{CI}_0(s)$ as the $s$ \emph{highest degree nodes}. This can be done as the two definitions are strongly correlated in the network models we focus on in this section \cite{fernholz2003giant, janson2007simple, luczak1991size}. For the uniform seeding, just as earlier, we choose the seed set $\mathcal{UI}_0(s)$ as $s$ nodes sampled uniformly at random in the whole network. In both setups we look at $\mathbb E_p[\mathbf{Cl}(\mathcal{CI}_0(s))] $ and $\mathbb E_p[\mathbf{Cl}(\mathcal{UI}_0(s))]$, the \emph{average percolation cluster sizes} of the initially infected nodes, when edges are retained with probability $p$. This corresponds to the average number of populations that experience local large outbreaks in the two seeding scenarios. The \emph{percolation pandemic size ratio function} is then defined as the ratio of these two averages:
\begin{equation}\label{eq:ratio}
    f_G(p,s)=\mathbb E_p[\mathbf{Cl}(\mathcal{CI}_0(s))]/\mathbb E_p[\mathbf{Cl}(\mathcal{UI}_0(s))],
\end{equation}
similarly to the earlier defined experimental function.

We define two approaches to the switchover phenomenon on a meta-network of $n$ cities. In the {\bf weak switchover phenomenon}, we require that there exists a seed count $s\le n$ and link-retention probabilities  $0<p_1, p_2 <1$ with
\begin{equation}\label{eq:switch}
 f_G(p_1,s)>1+c, \quad \text{and} \quad
f_G(p_2,s)<1-c,   
\end{equation}

for some constant $c$ that might depend on the network size.     Meanwhile, in the {\bf strong switchover phenomenon}, we require that the constant $c$ \emph{does not depend on the network size} $n$ and thus holds across a whole model class (e.g. GIRG or configuration model with fixed degree heterogeneity).  
When the switchover occurs for a seed count $s$, we say that the switch happens at retention probability  $p_{\mathrm{switch}}$ if $f_G(p,s)>1$ for $p<p_{\mathrm{switch}}$, while $f_G(p, s)<1$ for $p>p_{\mathrm{switch}}$. 

While the switchover phenomenon in GIRGs is hard to study analytically due to the lack of percolation theory developed for this  model, we borrow concepts from a simpler conventional network model, called Stochastic Block Model (SBM), to observe the strong switchover phenomenon. The SBM is able to mimic the central and rural areas of a population network, since it contains a `hidden geometry': We group populations into two sets of central or rural areas. Within areas we assume ageometric random networks, i.e., each pair of nodes is connected with the same probability, while the edge density between the two areas is lower.
\begin{theorem}\label{thm:Sto}
In the Stochastic Block Model with appropriately scaled parameters and $s_n=\Theta(n)$ the strong switchover phenomenon happens. (For a proof see the Supplementary Information (SI).)
\end{theorem}

In case of the ageometric configuration model, we are able to prove the weak switchover, already observed experimentally in Fig.~\ref{fig:3}c.
Further, we are able to  give quantitative bounds on $c$ in \eqref{eq:switch} as a function of the size $n$ of the population network $G$, the parameter $\tau$ expressing the prevalence of hubs, and the initial seed number $s=s_n$ that may also depend on the network size. 

\begin{theorem}\label{thm:config}
On the configuration model with exponent $\tau \in (2,4)$ and $1 \ll s_n \ll n$ the weak switchover phenomenon appears with $p_{\mathrm{switch}}$ slightly above the critical percolation parameter $p_c$. (For a proof see the SI.)
\end{theorem}

While Theorem \ref{thm:config} is valid for $\tau\in(2,4)$, the two regimes $\tau\in(2,3)$ and $\tau\in(3,4)$ quantitatively differ. In the former case, also called the scale-free regime, $p_c$ tends to zero as the network size grows and the region of the parameter space where seeding from nodes selected uniformly randomly is more dangerous is described by different linear equations compared to the $\tau\in(3,4)$ case. The switchover phenomenon disappears when $\tau>4$ as hubs become too small and separated from each other to produce the desired effect.

\begin{figure*}[t!]
    \centering
    \includegraphics[width = 1.\textwidth]{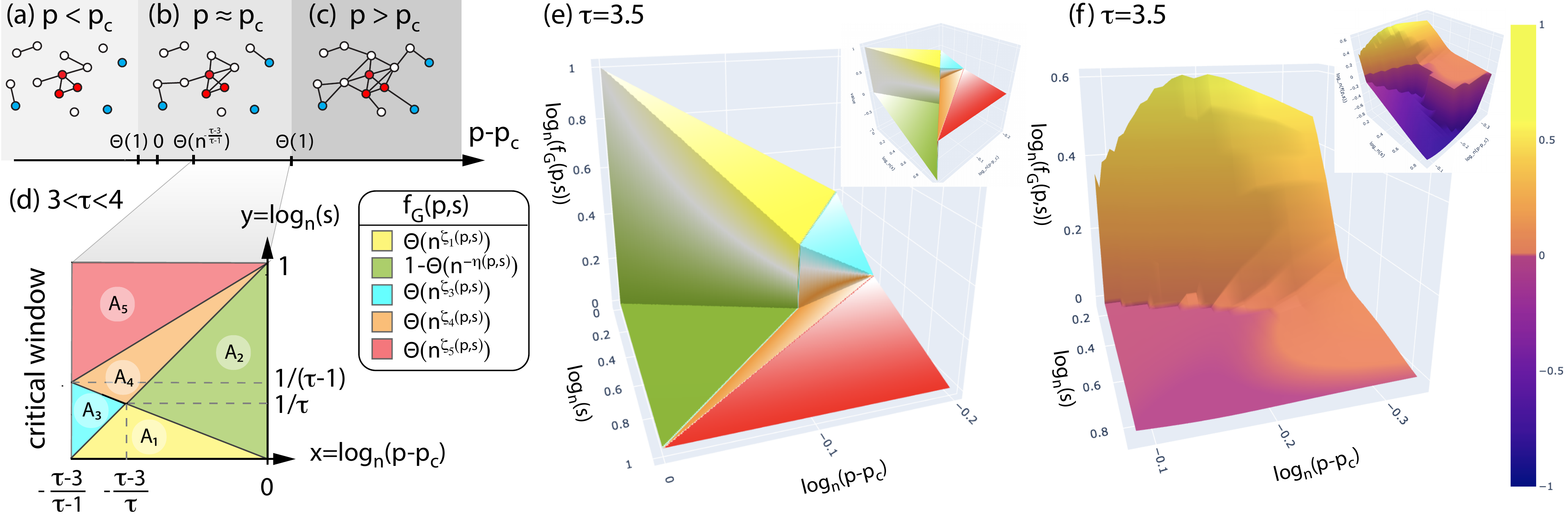}
    \caption{\small Panels (a), (b) and (c) show the heuristic explanation of the switchover of the pandemic size ratio function $f_G$. Panel (d) shows the phase diagram of the function $f_G$ for $3<\tau<4$, for values of $p$ slightly above the percolation threshold and for various values of $s$. The asymptotics of $f_G$ is different in parameter regions with different colors. The phase diagram for $2<\tau<3$ and the precise values of $\zeta_i$ in the legend of panel (d) are included in the Appendix. Panel (e) shows the 3D plot of $\bar{f}_{3.5}$, the limit function $\log_n(f_G)$ for $\tau=3.5$, as the number of nodes in $G$ tends to infinity, and panel (f) shows the corresponding simulation results on configuration model networks with $n=100000$ nodes (each datapoint is an average of 100 independent percolation instances on 10 indepedent random networks). The colors on panel (e) follows the colors on the phase diagram on panel (g). Since in the configuration model we only have weak switchover, the (green) part of the surface $\bar{f}_{3.5}$ that corresponds to $f_G<1$ converges to 0. For a visualization of the precise deviation of $f_G$ below 1, in the inset of panels (e) and (f), we plot the function $\log_n(f_G)$ for $f_G>1$, and $-\log_n(1-f_G)-1$ for $f_G<1$.}
    \label{fig:4}
\end{figure*} 
 
An interpretation of our rigorous derivations yield the following \emph{heuristic explanation} of the switchover phenomenon:
\begin{enumerate}
\item[a)] Below the percolation threshold, as demonstrated in Fig.~\ref{fig:4}a, the connected components of the central area seed nodes will be much larger than the components of the uniformly randomly selected seed nodes, however, they do not yet form a giant component. Nodes selected uniformly at random are not likely to be in these large components, hence the union of the connected components of seeds selected uniformly at random from the network will be smaller than the pandemic started from the central area. For very small values of $p$, seeding the highest degree nodes is the most dangerous on every graph.
\item[b)] Slightly above the percolation threshold, there will be a single ``giant component'' of  nodes experiencing a local pandemic, containing most of the central nodes. Thus when seeding starts from the central nodes, their components will contain this giant component and a small portion of smaller components. In the uniform seeding scenario, it is likely that a few of the random seed nodes will also belong to the giant component, while the other seeds, spread out randomly over the rest of the network, will be contained in lots of additional smaller components. Hence the union of the components of the uniformly chosen nodes will be larger, as shown in Fig.\!~\ref{fig:4}b.
\item[c)] Well above the percolation threshold (see Fig.\!~\ref{fig:4}c), there is essentially only one connected component thus each node gets infected regardless of the position of the seed nodes in the network.
\end{enumerate}

The phenomenon in b) is stronger when there are relatively few edges leaving the central area, which can be due to lack of long range interactions amplifying local geometric effects (as observed in GIRGs, see Fig.~\ref{fig:3}). However, in Theorem \ref{thm:Sto} the geometry induced by the two blocks is already enough to cause a strong switchover. On the contrary, there is nothing to limit the number of edges leaving the central area in the configuration model, (the degree-degree correlation coefficient is close to $0$ \cite{litvak2013uncovering, van2014degree,van2015degree}), hence the switchover phenomenon appears weak.
 
\subsubsection*{Quantitative results for the configuration model}

For geometric networks with various node degree distributions, \emph{critical exponents} have been already proposed earlier~\cite{newman2001random, cohen2002percolation}, with some of them proven rigorously for the configuration model (possibly with power-law degree distibution) \cite{dhara2021critical, dhara2016heavy, dhara2017critical, van2019component}, as well as for rank-1 inhomogeneous random graphs \cite{bhamidi2010scaling, bhamidi2012novel, van2018cluster} and for Erd\H{o}s-R\'enyi graphs \cite{ding2011anatomy}. Based on these results, we can prove that, after appropriate scaling, the pandemic size ratio $f_G$ of the configuration model (simulated on Fig.~\ref{fig:3}c for fixed $s$) converges to a two-dimensional limit function, which can be precisely determined. 
To state our result, let us re-parametrize $f_G(p,s)$ as a function of $x,y$ where $p=p_c+n^{x}$ for $x\in(-(\tau-3)/(\tau-1),0)$ and $s=s_n=n^{y}$ for $y\in(0,1)$, i.e., we consider $\widetilde f_G(x,y):=f_G(p_c+n^x, n^y)$. For $\tau\in(3,4)$, we divide the parameter space into five triangular regions $A_1$-$A_5$ illustrated on Fig.~\ref{fig:4}d (defined precisely in the SI). For $\tau\in(2,3)$, the picture is similar, except there is an extra triangular region $A_6$. 
\begin{theorem}
\label{thm:three}
On the configuration model with exponent $\tau\in(2,4)$,  $ \log_n(\widetilde f_G(x,y))$ converges to a function $\zeta(x,y)$. On each triangular region $A_i$, $\zeta(x,y)$ can be expressed as  a different linear function $\zeta_i(x,y)$. 
\end{theorem}
Theorem \ref{thm:three} implies that the percolation pandemic size ratio $f_G(p_c+n^x, n^y)=\Theta(n^{\zeta_i(x,y)})$ on $A_i$ for each $i=1$-$5$. We give the formula for each $\zeta_i$ in Methods, and the proof of Theorem \ref{thm:three} in the SI. A three-dimensional illustration of $\zeta$ can be seen in Fig.\!~\ref{fig:4}e. 
The limiting function $\zeta$ is discontinuous at the boundary line between $A_1$ and $A_2$ and between $A_3$ and $A_4$, respectively. These discontinuities correspond to a discontinuous phase transition of the system's behavior at those boundaries. Curves in Fig.~\ref{fig:3} correspond to horizontal cross-sections of the two-dimensional $f_G$ function for fixed $s$ values. We experience this phase transition on the curves of Fig \ref{fig:3} dropping steeply from above $1$ to below $1$ when slightly increasing $R_0$. Our result implies that the curves will look steeper and steeper as $n$, the size of the network increases. 

We also show that $f_G(p_c+n^x, n^y)= 1-\Theta(n^{\eta(x,y)})$ in region $A_2$. Hence, the scaling $\log_n (\widetilde f_G)$ in Theorem \ref{thm:three} is not appropriate in region $A_2$, which is reflected by the fact that the limiting function $\zeta_2$ is identically $0$ in this regime. To be able to compute how much $f_G$ (\eqref{eq:ratio}) goes below $1$ in $A_2$, i.e., how much more dangerous uniform seeding can be compared to central seeding, we extract the limiting exponent $\eta(x,y)$ using a different normalization for $f_G$, and give the formula in Methods. This different normalization is used on the area $A_2$ (green) in the inset of Fig.~\ref{fig:4}e, which in turn demonstrates that $f_G$ falls below $1$ in this regime. Finally, we validate our theoretical results for the configuration model by simulations in Fig.\!~\ref{fig:4}f. Despite the finite size of the simulations ($n=10^5$), the resemblance to the theoretical predictions is already apparent.

\section*{Discussion and Conclusions}

Different seeding of an epidemic can lead to significantly different outcomes depending on the actual value of the basic reproduction number $R_0$. While $R_0$ is defined by the biological parameters of the spreading disease, it is only one factor determining the effective reproduction rate $R_t$, which characterizes the actual speed of reproduction during an ongoing epidemic. In case of an influenza like disease, $R_t$ depends on many other factors including the actual number of interactions of people, the interventions at place, the self-protection measures (e.g. masks, sanitizing, etc.) or even the seasonal variance of temperature and humidity. Considering the distribution of the initial seeds and the actual $R_t$ values, our theory may suggests counter-intuitive effects during the consecutive waves of a pandemic. This could be the case in Hungary, where the first wave of the COVID-19 pandemic was initiated from large, well connected populations, while social distancing was very effective at the time, causing a smaller actual $R_t$ value during this period. Thus the clumped initial seeds and the relatively low $R_t$ could set relatively favorable conditions for the epidemic to reach a larger population, as compared to a uniformly seeded situation. Meanwhile, at the beginning of the second wave, seeded populations were more distributed all around the country, while social distancing was not followed rigorously. This induced larger $R_t$ values, which yet again set relatively easier conditions for the epidemic to reach a larger population, now seeded from a uniform initial state.

In this paper we studied the effects of epidemic seeding on geometric metapopulation networks. We were interested in the long-term behavior of spreading processes and showed that the relative danger of infecting a larger population when starting the process from the core or uniformly at random in a network has a non-monotonous dependency on $R_0$. We explored an entirely new switchover phenomenon and demonstrated them on real and synthetic networks via numerical simulations. We provided a rigorous proof for the existence of this phenomenon on a large set of random graphs, while we are confident that our theory can be extended for a more general set of graphs, which resembles certain structural constraints. Importantly, we identified the spatial geometry of the underlying structure as an important amplifying factor of the switchover phenomenon.

We build our theory on some results~\cite{newman2001random,cohen2002percolation}, which are broadly accepted by the network science community, yet it has not been proven rigorously for all network structure (for exceptions see SI). This implies certain limitations for our results, although assuming these results to hold, our proposed theory has been derived rigorously. In addition, we took some assumptions for the simplicity of our presentation but their generalization is possible. We demonstrated experimentally the switchover phenomenon on directed networks, while we assume an undirected structures in our theory, which can be extended for directed structures easily. Moreover, we conjecture that the observed phenomenon  occurs in most networks where a ``central'' region can be meaningfully distinguished in the structure. We concentrated on the conventional SIR model for the demonstration of the switchover phenomenon, but this observation holds for more realistic models, including the SEIR model with an addition compartment of exposed (E) state, better capturing the reaction scheme of the SARS Cov-2 disease. 

Beyond scientific merit, our results may contribute to the better designs of epidemic forecasts and intervention strategies in a country during an ongoing pandemic. We highlight the importance to follow not only the rate but also the spatial distribution of new infection cases of a spreading disease or its variants during the early phase of an epidemic. This could lead to new testing strategies, which disclose the spatial distribution of the epidemic during its initial phase, as this was the case in some countries (like  Denmark~\cite{holmager2020geography}) from the beginning of the COVID-19 pandemic. Based on these early-time observations our theory provides understanding about the long-term consequences of an epidemic by considering the commonly overlooked convoluted effects of epidemic seeding and the geometric structure of human populations and mobility.

\section*{Materials and Methods}

\subsection*{Data description}
\subsubsection*{Settlement level daily COVID-19 infection data for Hungary}

For the analysis presented in Fig.~\ref{fig:1} we used a dataset recording the daily number of newly infected cases in $3,118$ Hungarian settlements. This data matches the officially reported total number of daily cases~\cite{hungaryCOVID,hungaryCOVIDwiki}, however, just as the official data, it suffers from some observational bias due to the limited capacity of testing in the country during certain periods of the pandemic. For the analysis presented in Fig.~\ref{fig:1} we considered all settlements, and obtained their population sizes from data shared by the Hungarian Statistical Office~\cite{settlPopHun}. A version of this data aggregated on the county level is openly available~\cite{hungaryCOVIDwiki}.

\subsubsection*{Daily commuting network of Hungary}

For the data-driven simulations of the Hungarian epidemic we use a microcensus collected and released by the Hungarian Statistical Office in 2016~\cite{hungary2016census}. The data contains the number of people commuting for work or school on a daily base between the $3,186$ settlements in Hungary, with the districts of the capital considered as separate populations. In our analysis we concentrated only on settlements with populations larger than $1,000$ inhabitants and kept commuting links with at least $25$ daily commuters. From this data we constructed an undirected meta-population commuting network with $1,398$ settlements as nodes and $8,322$ commuting edges with weights computed as the average number of commuters between pairs of populations. The total population size of the network contained the $95 \%$ ($9,285,286$ individuals) of the Hungarian population. Despite the sparsity of the network ($0.85 \%$ of the possible edges are present), $19 \%$ of individuals commute between settlements on a daily base.

\subsection*{Moran's I statistic}
We compute the Moran's I statistic at time $t$ as
\begin{equation}
    I(t)=\frac{n\sum_{i,j} w_{ij}(y_i(t)-\bar{y}(t))(y_j(t)-\bar{y}(t))}{\sum_{i,j}w_{ij} \sum_i(y_i(t)-\bar{y}(t))^2},
\end{equation}
where $n$ is the number of nodes, $w_{ij}$ is the edge weight between the nodes $i$ and $j$, $y_i(t)$ is the number of new infected cases at  node $i$ at time $t$ and $\bar{y}(t)=(1/n)\cdot\sum_i y_i(t)$.

\subsection*{Generating Geometric Inhomogenous Random Graphs}
GIRG($\tau$, $\alpha$) networks were generated by the following process: the location of $n$ nodes are sampled uniformly at random from the square $[0,1]^2$, and each node $u$ is assigned with a ``fitness'' value ($w_u$) sampled from a power-law distribution with exponent $\tau$.
Each pair of nodes are connected by an edge with a probability
\begin{equation}\label{eq:girg-formula}
    P(u,v)=p \min  \left\{  \left(\frac{C w_u w_v}{n \| x_u-x_v\|^2 } \right)^\alpha, 1 \right\} ,
\end{equation}
which after only the largest connected component of the network is kept. To generate models with different parameters comparable to each other, we fix the number of edges to $5,000$, by selecting the constant $C$ and $p$ accordingly, since these two parameters are responsible for the edge-density. For the exact implementation see \cite{GIRG_code}. 
When the fitness distribution $w_u$ is set to be a power-law,  node degrees also satisfy a power law.
The abundance of long-range connections is tuned by $\alpha$ in \eqref{eq:girg-formula}: the smaller $\alpha$, the more likely are long-range connections (b).
The power-law exponent $\tau$ and the long-range parameter $\alpha$ tune the average graph distance in the network $\overline{\mathrm{Dist}}(n)$, see  \cite{deijfen2013scale, bringmann2016average, biskup2004scaling, deprez2015inhomogeneous}:
\begin{equation*}
    \label{eq:dist-girg}\overline{\mathrm{Dist}}(n)
= \begin{cases}
\Theta(\log \log n) & \text{when } \tau\in(2,3), \alpha>1\\
\Theta \big((\log n)^\zeta\big)  & \text{when } \tau>3, \alpha\in(1,2) \\
\Theta (\sqrt{n})  & \text{when } \tau>3, \alpha>2.
 \end{cases}
\end{equation*} 
Comparing this to the average distance in the configuration model, where only the first two regimes are possible ($\Theta(\log \log n)$ when $\tau\in(2,3)$, $\Theta(\log n)$ when $\tau>3$), and to distances in lattice models (where $\overline{\mathrm{Dist}}_N$ is polynomial in $n$), we observe that the underlying geometry of GIRGs with the long-range connections play a role when $\tau>3$, and the model interpolates between the small-world configuration model and the lattice.

\subsection*{Generating random networks from the Configuration Model}

We generate a uniform sample from the set of graphs with power-law degree distribution with degree exponent $\tau$ by first generating a GIRG with given parameter $\tau$ (and $\alpha=2.3$), and we swap the end-points of randomly selected pairs of edges~\cite{swapNetX} to remove all geometric and structural correlations from the structure, while conserving the degree of each node.. We perform $10 \cdot \mathrm{\#number\_of\_edges}$ swaps, which mixes the edges enough so that the resulting network becomes close to a uniform sample from the set of networks that have exactly the same degree sequence as the original GIRG network \cite{gkantsidis2003markov}. 

\subsection*{Core decomposition and seed selection}

In metapopulation network models, we use k-shell decomposition to identify the largest k-core of the network~\cite{carmi2007model,kshellNetX} and to select seeds in the central area. This algorithm computes the $k$-shell by recursively removing each node of the population network that has degree less than k, until no more nodes can be removed. We take the largest $k$ for which at least $s$ nodes remain, and we select $s$ nodes from this $k$-shell uniformly at random as our seed set in the central area. For the uniform seeding scenario, we select $s$ nodes of the population network uniformly at random. Finally, in both seeding strategies, we select $60$ agents in each of the $s$ nodes and we mark them as infectious agents at time 0.

In the theoretic computations and simulations, the $s$ highest degree nodes are selected for the central area, and $s$ uniformly random nodes are selected for the uniform seeding scenario. Node degrees and core-number of nodes in configuration network models are strongly correlated, allowing us to make this approximation.

\subsection*{SIR model on metapopulation networks}

We start by setting the home population  of $N_i$ agents to settlement $i$, where $N_i$ denotes the population of the actual settlement. Each agent is assigned exactly one home population, and the home assignments do not change for the rest of the simulation. We initialize the infection according to one of the seed selection scenarios and proceed with the simulation in each iteration $t$ in three steps. In the diffusion step, each agent who is at its home population $i$ selects a target population $j$ with probability $p_mw_{ij}/N_i$ and moves there. Agents that are not at their home population simply move back to their home settlement. In the reaction step, each susceptible agent in population $i$ becomes infected with probability $1-(1-\beta/N_i)^{I_i}$, where $I_i$ is the number of infected agents in population $i$ at iteration step $t$ and $\beta$ is the infection rate. In the final recovery step,  each infected agent recovers with rate $\mu$.

\subsection*{The limiting function of the percolation pandemic size ratio}
In Theorem \ref{thm:three}, we identified the scaling of $f_G(p, s)=f_G(p_c+n^x, n^y)=\Theta(n^\zeta(x,y))$, where $\zeta(x,y)$ is a piecewise linear function. On each region $A_1$-$A_6$, $\zeta$ is given as follows:  
\begin{equation*}
\zeta(x,y) =
\begin{cases}
\zeta_1(x,y)=1+\left(\frac{1}{|\tau-3|}+\mathbbm{1}_{\tau \in (3,4)}\right)x-y &\mbox{on } A_1 \\
\zeta_2(x,y)=0 & \mbox{on } A_2 \\
\zeta_3(x,y)=\mathbbm{1}_{\tau \in (3,4)}x+\left(1-\tfrac{1}{\tau-1}\right)(1-y) & \mbox{on } A_3 \cup A_6 \\
\zeta_4(x,y)=-\tfrac{1}{|\tau-3|} x-\tfrac{1}{\tau-1}(1-y) &\mbox{on } A_4 \\
\zeta_5(x,y)=\frac{1}{(\tau-1)(\tau-2)}(1-y) &\mbox{on } A_5
\end{cases}
\end{equation*}
Finally, on region $A_2$, $f_G(p_c+n^x, n^y)=1-\Theta(n^{-\eta(x,y)})$ where 
\begin{equation*}
 \eta(x,y)=\left(\frac{1}{|\tau-3|}+\mathbbm{1}_{\tau \in (3,4)}\right)x-y.   
\end{equation*}

\section*{Acknowledgements}
We are grateful for the insightful comments by Mikl\'os Abert, G\'eza \'Odor, and Tim Hulshof and for the shared datasets by J\'anos K\"oll\H{o}, P\'eter Pollner, Mikl\'os Sz\'ocska, Beatrix Oroszi, Gergely R\"ost and the COVID-19 Force Group of Hungary. This work has been supported by the Dynasnet ERC Synergy project (ERC-2018-SYG 810115). G\'O acknowledges support from the Swiss National Science Foundation (200021-182407), DC received support from the grant MILAB 2018-1.2.1-NKP-00008, and MK was supported by the DataRedux ANR project (ANR-19-CE46-0008) and the SoBigData++ H2020 project (H2020-871042).

\section*{Author contributions statement}
All authors designed the research. G\'O, JK, and LL performed the analytical calculations, DC, G\'O and MK designed the numerical simulations, and MK and G\'O carried out the data analysis. All authors wrote the final manuscript. 

\newpage

\setcounter{table}{0}
\renewcommand{\thetable}{S\arabic{table}}%
\setcounter{figure}{0}
\renewcommand{\thefigure}{S\arabic{figure}}%
\setcounter{section}{0}
\renewcommand{\thesection}{S\arabic{section}}%

\hspace{-.2in}{\huge\textbf{Supplementary Information}} \\ \\


We state precise versions of Theorems 1,2 and 3 of the main text. 

\section{Strong switchover}
\label{sec:strong}

We show that there exist graph sequences with the strong switchover property.

Let $V$ be a set of $n$ nodes, partitioned as $V=U\cup W$, where $U$ is the ``central region'' and $W$ is the ``periphery''; let  
$|U|=|W|=n/2$. We construct a stochastic block type random graph $G$ on $V$ as follows: within $U$, we connect any two nodes with probability $a/n$; we connect any other pair with probability $b/n$, where $a$ and $b$ are sufficiently large constants. It will turn out that only the ratio $a/b$ matters; we assume that $a/b>200$. Let's assume that there is an epidemic spreading on this graph and $\beta$ is the probability that an edge passes on the infection, if one endpoint is infected and the other one is not. Let  $G^p$ be the subgraph of $G$ obtained by deleting every edge with probability  $1-p$.

In Experiment 1, we infect a seed $S_1$ of $s<n/2$ random central nodes (denoted by $\mathcal{CI}_0(s)$ in the main text); in Experiment 2, we infect a seed $S_2$ of $s$ random nodes from the whole graphs (denoted by $\mathcal{UI}_0(s)$ in the main text). Let  $f_i=|G^p(S_i)|$ be the number of removed nodes at the end of Experiment $i$. We are interested in the ratio $f_G(p,s)=\E_p[f_1]/\E_p[f_2]$. 

\begin{theorem}\label{thm:baby}
With high probability as $n\to\infty$, for sufficiently small values of  $p$, $\E_p[f_1]>(3/2)\E_p[f_2]$; for sufficiently large values of $p$, $\E_p[f_2]<(11/12)\E_p[f_1]$. If  $p\to 1$  then $\E_p[f_2]\sim\E_p[f_1]$.
\end{theorem}

\subsection{Proof of strong switchover results}

To prove Theorem \ref{thm:baby}, we prove some  more general facts about random graphs. Let $V=\{1,\dots,n\}$ and let $Q=(q_{ij})_{i,j=1}^n$ be a symmetric matrix with $0\le q_{ij}\le 1$. Let $G=G(P)$ be an inhomogeneous random graph $V$, obtained by connecting nodes $i$ and $j$ with probability $q_{ij}$. Let $S\subseteq V$, $|S|=m$, and let $G(S)$ be the union of connected components of $G$ intersecting $S\subseteq V$ (in the main text, the cardinality for this set is denoted by $\mathbf{Cl}(S)$ without specifying the underlying graph).

\begin{lemma}\label{LEM:GBU}
Let $0<\gamma<1$. {\rm(a)} If $\sum_{j\in V\setminus S} q_{ij}\le \gamma$ for
all nodes $i\in V$, then
\[
\E[|G(S)|] \le \frac1{1-\gamma}m.
\]
{\rm(b)} If $q_{ij}=\gamma/n$ for all $i\not= j$, $i\in V\setminus S$, then
\[
\E[|G(S)|] \ge \frac{n+\gamma n}{n+\gamma m}\,m.
\]
\end{lemma}

\begin{proof} Throughout this proof, we denote the vertex set of of a graph $H$ by $V(H)$, and write $|H|$ for the number of nodes in $V(H)$.
We construct the connected component of $G$ containing a node $u$ by a simultaneously performing a   depth-first search (DFS) exploration that we describe below and deciding about whether there is an edge between each pair of nodes as we encounter them in this DFS. We start the exploration with the edge-less forest $F_0$ on $V(F_0)=S$ containing the seed nodes. Suppose that a forest $F_k$ with $|F_k|\ge k$ has been already constructed, and $k$ of its nodes have been ``scanned'' (equivalently, explored), meaning that the status of all edges incident to them has been decided. If all nodes of $F_k$ have been scanned, we stop. Else, let $v$ be any unscanned node of $F_k$. We add each edge between $v$ and $w\in V\setminus V(F_k)$ to $F_k$ with probability $q_{vw}$, and obtain so $F_{k+1}$. Note that in case we do not happen to add any new vertices to $F_k$, then $V(F_k)=V(F_{k+1})$. At the exploration step we also decide about the edges between $v$ and the unscanned nodes of $F_k$, but these edges play no role in increasing the size of $G(S)$. We label $v$ as scanned, and move to the next unscanned node in $F_{k+1}$.

If we stop after $\tau$ steps, then $V(F_\tau)= V(G(S))$.
 Since every node of $F_\tau$ has been scanned exactly once, we have $|F_\tau|=|G(S)|=\tau$. For $k<\tau$, we calculate the expected size of $F_{k+1}$ by using the expected number of edges added in the $k$-th exploration step:
\[
\E[|F_{k+1}|\mid F_k]= |F_k| + \sum_{j\in V\setminus V(F_k)} q_{vj}
\le |F_k| + \gamma,
\]
where we used the assumption of the lemma to obtain the last inequality.
Hence the random variables $Y_k=|F_k|-\gamma k$ form a supermartingale, and so
\[
\E[Y_0] = |S|=m \ge \E[Y_\tau] = \E[|G(S)|] - \gamma \E[\tau] = (1-\gamma)\E[|G(S)|].
\]
This proves the first inequality. We  prove the second inequality by counting nodes reachable via a single edge from $S$.  The probability that $u\notin S$ is not connected to any node of $S$ is
\[
\mathbb P(u \nleftrightarrow S) = \Big(1-\frac{\gamma}{n}\Big)^m \le  e^{-\gamma m/n} \le \frac1{1+\gamma m/n},
\]
and so $u$ is connected to $S$ with probability at least
\begin{equation}\label{eq:connect-prob}
    \mathbb P(u \leftrightarrow S) \ge 1-\frac1{1+\gamma m/n} = \frac{\gamma m}{n+\gamma m}.
\end{equation}

Since expectation is linear, $\E[|G(S)|] \ge |S| + \sum_{u\in V\setminus S} \mathbb P(u \leftrightarrow S)$. Using now \eqref{eq:connect-prob} we get the inequality in part (b) of the Lemma.
\end{proof}

Turning to the proof of Theorem \ref{thm:baby}, first we treat the case when $p$ is small. Recall that we form $G$ as decomposing its vertex set as $V=U\cup W$, with $U$ standing for the central region and having connection probability $a/n$ between its vertices, while every other pair of vertices has connection probability $b/n$.  For any subset $S\subseteq V$ with $|S|=s$ and $p\to0$,   
\[
\E[|G^p(S)|] = s +p\sum_{v\in S}\deg(v) + O(p^2),
\]
since the probability of every particular infection path of length 2 or more is at most $p^2$, and the probability that a node is counted twice is $O(p^2)$. So it follows that for a very small $p$, larger total degree of the seed set implies that the final size of the epidemics is larger. A bit more careful computation, using Lemma \ref{LEM:GBU}(a) shows that in the case of our graph, seeding $s$ random nodes in the central set $U$ is more dangerous than seeding $s$ random nodes in the whole underlying set $V$ for $p<1/a$.

Next, consider the case of a larger $p$, say $p=1/(4b)$. Let $s=n/3$ (any $s<n/2$, $s=\Omega(n)$ would do). Due to the independence of keeping edges in the bond-percolation, we observe that the percolated random graph $G^p$ is an inhomogeneous random graph itself, with edge probabilities $p_{ij}=ap/n$ when both $i,j\in U$ and $p_{ij}=bp/n$ otherwise. 
With high probability, the subgraph $G^p[U]$ (that is the subgraph of $G^p$ restricted to nodes in $U$) will have a giant component $H_0:=\mathcal C_{\max}(G^p[U])$ of size 
\begin{equation}\label{EQ:H0-LARGE}
|V(H_0)| \ge (1-e^{-p a/2})|U| > \Big(\frac12-\frac1{2^{17}}\Big) \frac{n}2
\end{equation}
(see \cite{janson2011random}, Theorem 5.4). The subgraph $H_0$ extends to a component $H$ of $G$, which may contain additional nodes from $W$, and through this, even some additional nodes of $U$. 

Let $|U\setminus V(H)|=\mu n$ and $|W\setminus V(H)| = \nu n$. By \eqref{EQ:H0-LARGE} $\mu<2^{-17}$. Applying Lemma \ref{LEM:GBU}(a) with $S=U$ and with $\gamma=pb/n=1/4$ (note that $H_0$ is defined solely from the edges of $G[U]$, and to apply the lemma, we only need the randomness of the edges in $E(G^p)\setminus E(G[U])$; the edge probabilities within $U$ play no role) we get that 
\[
\E[|V(H)\cap W|~\big|~H_0] = \frac{n}{2}-\E[\nu] n \le \frac{\gamma}{1-\gamma}\,\frac{n}{2} = \frac{n}{6}.
\]
Similarly, we can apply Lemma \ref{LEM:GBU}(b) to the graph $G^p[V(H_0)\cup W]$ and $S=V(H_0)$, to get 
\[
\E[|V(H)\cap W|~\big|~H_0] > \frac{n}{18}.
\]
So $1/3 <\nu< 4/9$ in expectation over $H$ for almost all $H_0$.

We condition on $H$, which means to freeze the edges and non-edges incident with $H$, so we still have an inhomogeneous random graph $G'=G^p[V\setminus V(H)]$. The probability that \eqref{EQ:H0-LARGE} is violated is negligible, so we assume that $H_0$ satisfies it. The expected degree of a node $v$ in $G'$ is $(\mu+\nu)n(\gamma/n) = (\mu+\nu)/4$ if $v\in V(G')\cap W$, and $\mu n (p a/n)+ \nu n (p b/n) = 50\mu+\nu/4$ if $v\in V(G')\cap U$. In both cases, this is bounded by $1/3$. 

In Experiment 1, picking $s$ random nodes in $U$, we meet $H$ with high probability, and we pick some $s_1\le \mu n$ nodes in $U\setminus V(H)$. Clearly $\E(s_1)=2\mu s$. By Lemma \ref{LEM:GBU}(a), these nodes infect at most $3\mu s$ nodes of $G'$ in expectation, and so
\begin{align}\label{EQ:U-SEED2}
\E\big[|G^p(S_1)|~\big|~H\big] &= |V(H)|+ \E[|G^p(S)\setminus V(H)|]  \le |V(H)| + 3\mu s. 
\end{align}
In Experiment 2, picking $s$ random nodes in $V$, with high probability we hit $H$, and an expected number of $(\mu+\nu)s$ nodes of nodes in $V\setminus V(H)$. Even ignoring further nodes infected by these nodes, we get 
\begin{equation}\label{EQ:U-SEED3}
\E\big[|G^p(S_2)|~\big|~H\big] \ge |V(H)|+ \frac{(n-|V(H)|)s}{n} =  |V(H)|+ (\mu+\nu)s. 
\end{equation}
Subtracting \eqref{EQ:U-SEED2} from \eqref{EQ:U-SEED3} yields 
\[ 
\E\big[|G^p(S_2)|~\big|~H\big] - \E\big[|G^p(S_1)|~\big|~H\big] \ge (\nu-2\mu)s.
\]
Taking expectation over $H$, we get that
\[ 
\E\big[|G^p(S_2)|\big] - \E\big[|G^p(S_1)|\big] \ge \frac14s.
\]
Hence, since we took $s=n/3$,
\[
\frac{\E[|G^p(S_1)|]}{\E[|G^p(S_2)|]} \le 1-\frac{s}{4n} = \frac{11}{12}.
\]
This proves that strong switchover occurs.

Finally, it is clear that if $p$ is very close to $1$, then $\E(R_1),\E(R_2)\sim n$.

\section{Weak switchover}
\label{sec:weak}

In this subsection we focus on the Configuration model. We start by introducing new notation needed for stating the results and the proofs. See Table \ref{tab:glossary} for a glossary of notations.

\begin{definition}[Configuration model and its percolation]
Let us denote the Configuration model on $n$ nodes and degree exponent $\tau$ by $\mathrm{CM}(n,\tau)$. Let $G^{p}_n$ be the percolated $\mathrm{CM}(n,\tau)$ with edge-retention probability $p$ on $n$ vertices. Denote by $n_c$ be the number of connected components of $G^{p}_n$, by $\mathcal{C}_i$ the $i^{th}$ largest component of $G^{p}_n$, by $\mathcal{C}(u)$ the component in $G^{p}_n$ which contains node $u$, and let $C_i=\E[|\mathcal{C}_i|]$. 
Let $p_{c,n,\tau}$ be the critical percolation parameter for $\mathrm{CM}(n,\tau)$ for the existence of a linear sized giant connected component. For edge-retention probability $p=p_n$ that may depend on $n$, we define $\theta_n=p_n-p_{c,n, \tau}$. 
\end{definition}

Recall that $\mathbb E_p[\mathbf{Cl}(\mathcal{CI}_0(s))]$ is the expected cluster size of the seed set with the $s$ highest degree nodes in the percolated graph with retention probability $p$, and $\mathbb E_p[\mathbf{Cl}(\mathcal{UI}_0(s))]$ is the same for the seed set with $s$ uniformly chosen nodes. We are interested in the function $f_G(p,s)$, which is the ratio of these two expectations. In particular, we prove the week and the strong switchover properties in terms of the function $f_G(p,s)$ as defined in the main text. Now we are ready to state the precise version of Theorem 1 there.

\begin{theorem}
\label{thm:weak}
The sequence of random graphs sampled from the Configuration model with exponent $\tau \in (2,4)$ and $n \rightarrow \infty$ exhibit weak switchover. Specifically, under the assumptions $1 \gg \theta_n \gg n^{-|\tau-3|/(\tau-1)}$, and $n \gg s_n \gg 1$,
\begin{enumerate}[leftmargin=1cm]
\item if $\theta_n^{-\ptaumTh}  \gg s_n$ or  $ s_n \gg n \theta^{\frac{\tau-1}{|\tau-3|}}$, then $f_{\mathrm{CM}(n,\tau)}(p_{c,n,\tau}+\theta_n,s_n) > 1$,
\item if $\theta_n^{-\ptaumTh}  \ll s_n \ll n \theta_n^{\frac{\tau-1}{|\tau-3|}}$, then $f_{\mathrm{CM}(n,\tau)}(p_{c,n,\tau}+\theta_n,s_n)<1$,
\end{enumerate}
with high probability as $n\to\infty$.
\end{theorem}

Theorem \ref{thm:weak} is a qualitative result that shows the existence of the weak switchover phenomenon. In Theorem \ref{thm:weak_quan} we report our quantitative results on $f_{\mathrm{CM}(n,\tau)}(p,s)$, which will directly imply Theorem \ref{thm:weak}. First we need some new definitions.

\begin{definition}[Phases of the parameter space]
\label{def:Ai}
Given $\tau \in (2,4)$ and a sequence $(\theta_n)_{n\ge 1}=(p_n-p_{c, n, \tau})_{n\ge 1}$ with $(\theta_n>0)$, and  $(s_n)_{n\ge 1}$ with $(s_n>0)$ for which the limits $x = \lim_{n \rightarrow \infty} \log_n(\theta_n)$ and $y = \lim_{n \rightarrow \infty} \log_n(s_n)$ both exist, let us partition the parameter space $(\theta_n, s_n)$ (equivalently, $(x, y)$) into six sets in the following way

\begin{align*}
A_1&= \big\{ (\theta_n, s_n)\mid s_n \ll \mathrm{min}(\theta_n^{-\ptaumTh}, n \theta_n^{\frac{\tau-1}{|\tau-3|}})  \big\}  \equiv\big\{(x, y) \mid y < \min\{ -\tfrac{1}{|\tau-3|} x, 1+ \tfrac{\tau-1}{|\tau-3|} x\} \big\} \\
A_2&= \big\{ (\theta_n, s_n) \mid \theta_n^{-\ptaumTh}  \ll s_n \ll n \theta_n^{\frac{\tau-1}{|\tau-3|}}  \big\} \equiv\big\{(x, y) \mid -\tfrac{1}{|\tau-3|} x < y < 1+ \tfrac{\tau-1}{|\tau-3|} x \big\} \\
A_3&=\big\{ (\theta_n, s_n)\mid n \theta_n^{\frac{\tau-1}{|\tau-3|}} \ll s_n \ll  \theta_n^{-\ptaumTh} \big\} \equiv\big\{(x, y) \mid 1+ \tfrac{\tau-1}{|\tau-3|} x \le y \le  -\tfrac{1}{|\tau-3|} x \big\} \\
A_4&= \big\{ (\theta_n, s_n) \mid  \mathrm{max}(\theta_n^{-\ptaumTh}, n \theta_n^{\frac{\tau-1}{|\tau-3|}}) \ll s_n \ll  
\mathrm{min}(n \theta_n^{\frac{\tau-2}{|\tau-3|}},
n \theta_n^{\frac{1}{|\tau-3|}})\big\} \\& \equiv\big\{ (x, y) \mid \max\{  -\tfrac{1}{|\tau-3|} x, 1+ \tfrac{\tau-1}{|\tau-3|} x \} < y < \min\{ 1+ \tfrac{\tau-2}{|\tau-3|}x, 1+ \tfrac{1}{|\tau-3|}x \} \big\}\\
A_5&= \big\{ (\theta_n, s_n) \mid  n \theta_n^{\frac{\tau-2}{|\tau-3|}} \ll s_n  \big\} \equiv\big\{(x, y) \mid 1+ \tfrac{\tau-2}{|\tau-3|}x < y  \big\} \\
A_6&= \big\{ (\theta_n, s_n) \mid  \mathrm{max}(\theta_n^{-\ptaumTh}, n \theta_n^{\frac{1}{|\tau-3|}}) \ll s_n \ll  
n \theta_n^{\frac{\tau-2}{|\tau-3|}}) \big\} \equiv\big\{(x, y) \mid -\tfrac{1}{|\tau-3|} x < y < 1+ \tfrac{\tau-2}{|\tau-3|}x\big\}
\end{align*}

\end{definition}

\begin{figure}[ht!]
\begin{center}
  \includegraphics[width=\textwidth]{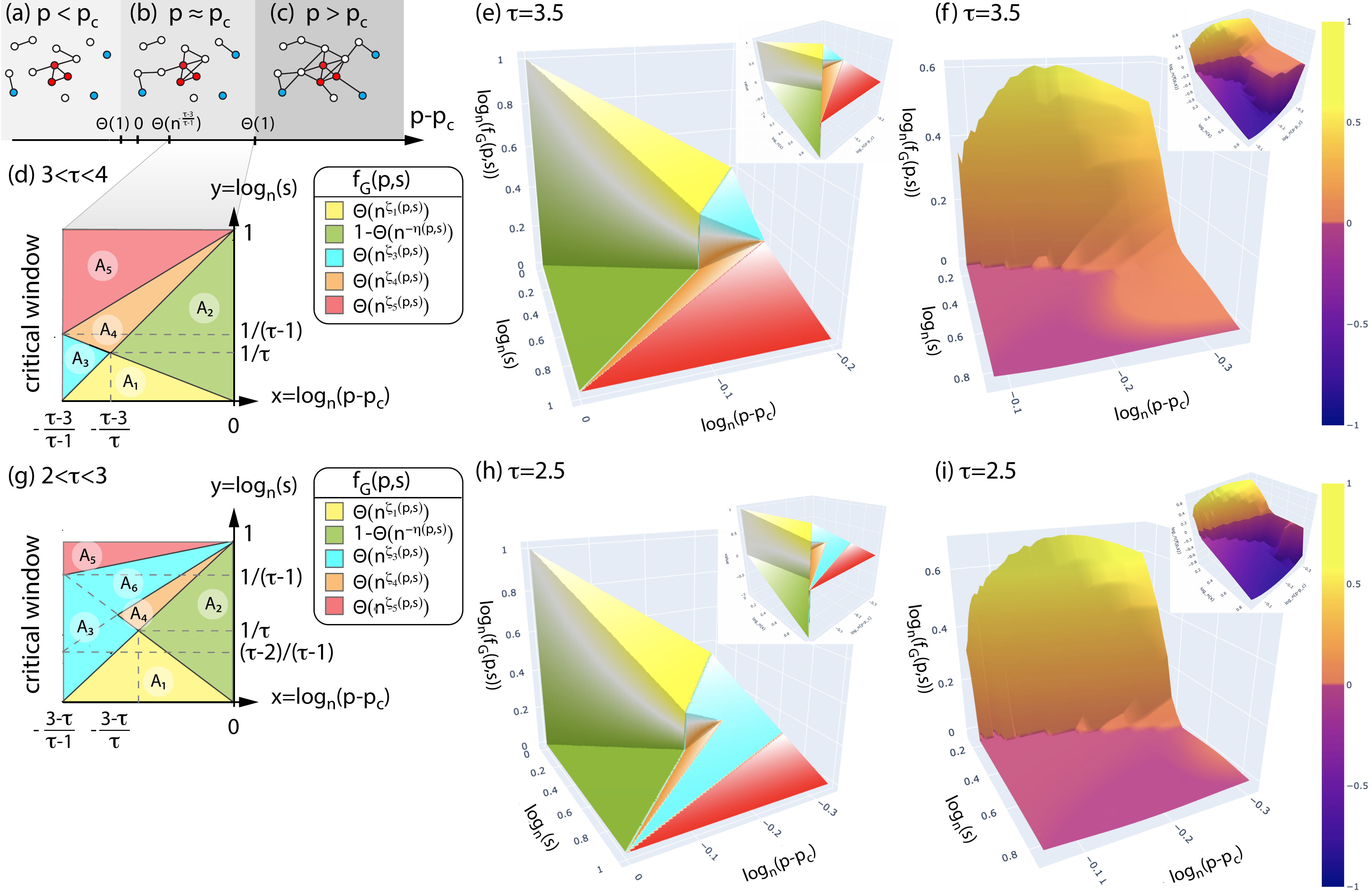}
  \caption{\small Subfigures (a), (b) and (c) show the heuristic explanation of the switchover of the pandemic size ratio function $f_G$. Subfigures (d) and (g) show the phase diagram of the function $f_G$ for $3<\tau<4$ and $2<\tau<3$, respectively, for values of $p$ slightly above the percolation threshold and for various values of $s$. The asymptotics of $f_G$ is different in the differenly colored parameter regions, which correspond to $A_1$-$A_6$ as given in Definition \ref{def:Ai}. Subfigures (e) and (h) show the 3D plot of $\bar{f}_{3.5}$ and $\bar{f}_{2.5}$, the limit function $\log_n(f_G)$ for $\tau=3.5$ and $\tau=2.5$, respectively, as the number of nodes in $G$ tends to infinity, and subfigures (f) and (i) show the corresponding simulation results on configuration model networks with $n=100,000$ nodes (each datapoint is an average of 100 independent percolation instances). The colors on subfigures (e)-(i) follow the colors on the phase diagram on subfigure (g). Since in the configuration model we only have weak switchover, the (green) part of the surface $\bar{f}_{\tau}$, which corresponds to $f_G<1$, converges to 0. For a visualization of the precise deviation of $f_G$ below 1, in the inset of subfigures (e) (f), (h) and (i) we plot the function $\tilde{f}_{\tau}$ and $\mathrm{dNorm}(f_G)$.}
  \label{result_ranges}
  \end{center}
\end{figure}

See Figure \ref{result_ranges} for a visualization of the sets $A_i$. Intuitively, the union of $A_1$ and $A_3$ are the parameter ranges for which  there is no uniformly selected seed in the giant. The union of $A_1$ and $A_2$ are the parameter ranges for which all of the high degree seeds are contained in the giant. The set $A_4$ is an intermediate regime, where there are high degree seeds outside the giant and uniform seeds inside the giant, and in $A_5$ the parameter $s$ is so large that there are multiple uniformly selected seeds in medium sized components (in addition to the giant). We note that $A_6$ is an empty set for $\tau \in (3,4)$. For $\tau \in (2,3)$, the set $A_6$ contains the parameter ranges where the giant component is smaller than the contribution of small components with only a single uniformly selected seed.

For our quantitative results, we will have to condition on an event $\mathcal{E}_n$, which holds with high probability. This is a standard technique to rule out rare events that have too big of an impact on the expected value (see the analytic derivation for more details). In the next definition we extend the definition of $f_G(p,s)$ to incorporate this conditioning.

\begin{definition}[Percolation pandemic size ratio]
On a graph $G$ and two seeding sets $\mathcal CI_0(s), \mathcal CU_0(s)$ of size $s$, with edge-retention probability $p\in[0,1]$, let the pandemic size ratio function conditioned on an event $\mathcal{E}$ be
\begin{equation}\label{eq:ratio}
    f_G(p,s,\mathcal{E})=\frac{\mathbb E_p[\mathbf{Cl}(\mathcal{CI}_0(s)) \mid \mathcal{E}]}{\mathbb E_p[\mathbf{Cl}(\mathcal{UI}_0(s)) \mid \mathcal{E}]}.
\end{equation}
 
\end{definition}

Now we are ready to state the precise version of Theorem 3 from the main text.

\begin{theorem}
\label{thm:weak_quan}
Let $A_i$ defined in Definition \ref{def:Ai}, for a sequence random graphs sampled from the Configuration model with exponent $\tau \in (2,4)$, and let $\mathcal{E}_n$ be the event that either $ s_n\in A_2 \cup A_4 \cup A_5 \cup A_6$ or the event $\{ \mathcal{UI}_0(s_n) \cap \mathcal{C}_1 = \emptyset,  s_n\in A_1 \cup A_3 \} $ hold. Then, if $p_{c,n,\tau}$ is the critical percolation parameter for $\mathrm{CM}(n,\tau)$, under assumptions $1 \gg \theta_n \gg n^{-(|\tau-3|)/(\tau-1)}$, and $n \gg s_n \gg 1$,
\begin{equation}
f_{\mathrm{CM}(n,\tau)}(p_{c,n,\tau}+\theta_n,s_n,\mathcal{E}_n)=
\begin{cases}
\Theta(\theta_n^{\frac{1}{|\tau-3|}+\mathbbm{1}_{\tau \in (3,4)}}n/s_n) &\mbox{if } (s_n, \theta_n) \in A_1 \\
1-\Theta(\theta_n^{-\frac{1}{|\tau-3|}-\mathbbm{1}_{\tau \in (3,4)}}s_n/n) & \mbox{if } (s_n, \theta_n) \in A_2 \\
\Theta(\theta_n^{\mathbbm{1}_{\tau \in (3,4)}}(n/s_n)^{1-\ptaumOn}) & \mbox{if } (s_n, \theta_n) \in A_3 \cup A_6 \\
\Theta( \theta_n^{-\ptaumTh}(n/s_n)^{-\ptaumOn}) &\mbox{if } (s_n, \theta_n) \in A_4 \\
\Theta((n/s_n)^{\frac{1}{(\tau-1)(\tau-2)}}) &\mbox{if } (s_n, \theta_n) \in A_5
\end{cases},
\end{equation}
and $\P(\mathcal{E}_n) \rightarrow 1$.
\end{theorem}

Since for $A_1, A_3, A_4, A_5, A_6$ we have $f_{\mathrm{CM}(n,\tau)}(p_n,s_n,\mathcal{E}_n) \rightarrow \infty$ and for $A_2$ we have $f_{\mathrm{CM}(n,\tau)}(p_n,s_n,\mathcal{E}_n) \rightarrow 1$, some of our results will be lost if we apply the same normalization to the limit of $f_{\mathrm{CM}(n,\tau)}(p_n,s_n,\mathcal{E}_n)$ for all $A_i$. For example if we normalize by applying the function $\log_n$, as we do in the definition of $\bar{f}_\tau$ in Definition \ref{def:Rtilde}, the deviation of $f_G$ below 1 in the region $A_2$ will disappear. To mitigate this issue, we propose a discontinuous normalization in addition to normalizing by $\log_n$.
\begin{definition}
\label{def:Rtilde}
Let $\theta=(\theta_n)=(p_n-p_{c,n,\tau})$, $s=(s_n)$ be a sequence of seed counts, $\mathcal{E}=(\mathcal{E}_n)$ be a sequence of events and for $z>0$ let us define the normalisation

\begin{equation}
\mathrm{dNorm}_n(z) =\begin{cases}
 \log_n(z) &\mbox{if } z>1 \\
 -\log_n(1-z)-1 & \mbox{if } z<1
\end{cases}.
\end{equation}
Then, assuming the limits $x = \lim_{\rightarrow \infty} \log_n(\theta_n)$ and $y = \lim_{n \rightarrow \infty} \log_n(s_n)$ both exist, we define
\begin{align*}
\bar{f}_\tau(x,y,\mathcal{E}) &= \lim\limits_{n \rightarrow \infty} \log_n(f_{\mathrm{CM}(n,\tau)}(p_{c,n,\tau}+\theta_n,s_n,\mathcal{E}_n)) \\
\tilde{f}_\tau(x,y,\mathcal{E}) &= \lim\limits_{n \rightarrow \infty} \mathrm{dNorm}_n(f_{\mathrm{CM}(n,\tau)}(p_{c,n,\tau}+\theta_n,s_n,\mathcal{E}_n)).
\end{align*}
\end{definition}

Now we are ready to apply the normalization and find the limiting curve. See Figure \ref{result_ranges} for a visualization in 3D.

\begin{corollary}
\label{cor:weak_quan}
Let $A_i$ and $\mathcal{E}$ be defined as in Theorem \ref{thm:weak_quan}, and $x, y, \tilde{f}_\tau(x,y,\mathcal{E})$ as given in Definition \ref{def:Rtilde}. Then, under the assumptions $0 > x > -(|\tau-3|)/(\tau-1)$, and $1 > y > 0$,
\begin{equation}
\tilde{f}_\tau(x,y,\mathcal{E}) =
\begin{cases}
1+\left(\frac{1}{|\tau-3|}+\mathbbm{1}_{\tau \in (3,4)}\right)x-y &\mbox{if } (x, y) \in A_1 \\
\left(\frac{1}{|\tau-3|}+\mathbbm{1}_{\tau \in (3,4)}\right)x-y & \mbox{if } (x, y) \in A_2 \\
\mathbbm{1}_{\tau \in (3,4)}x+\left(1-\ptaumOn\right)(1-y) & \mbox{if } (x, y) \in A_3 \cup A_6 \\
-\ptaumTh x-\ptaumOn(1-y) &\mbox{if } (x, y) \in A_4 \\
\frac{1}{(\tau-1)(\tau-2)}(1-y) &\mbox{if } (x, y) \in A_5
\end{cases},
\end{equation}
and $\P(\mathcal{E}_n) \rightarrow 1$. Moreover $\bar{f}_\tau(x,y,\mathcal{E})=\tilde{f}_\tau(x,y,\mathcal{E})$ except if $(x, y) \in A_2$, when $\bar{f}_\tau(x,y,\mathcal{E})=0$.
\end{corollary}

\begin{remark}
As shown in Figure \ref{result_ranges} (e) and (h), with the continuous normalizaton $\bar{f}$, there is a continuous transition between all of the regions except on the boundary of regions $A_1$ and $A_2$ and on the boundary of regions $A_3$ and $A_4$, where the transition is discontinuous.
\end{remark}

\subsection{Proof of weak switchover results}

\subsubsection{Previous results on random networks with power-law degree distribution}
\label{sec:prev_res}
Percolation cluster sizes in the near-critical regime has been extensively studied in the physics literature for various network models. For a-geometric networks with power-law degree distributions, the non-rigorous works \cite{newman2001random, cohen2002percolation} predict \emph{critical exponents}, some of it has been made rigorous for the configuration model \cite{dhara2021critical, dhara2016heavy, dhara2017critical, van2019component} for rank-1 inhomogeneous random graphs \cite{bhamidi2010scaling, bhamidi2012novel, van2018cluster} and Erd\H{o}s-R\'enyi graphs \cite{ding2011anatomy}. Based on these results, we summarize the cluster size distribution in the near-critical regime in Table \ref{tab:SF1}, and in a reparametrized form in Table \ref{tab:SF2}. To unify the notation we denote by $\theta_n$ the deviation of $p_n$ from the critical point (denoted by $s^\star$ in the physics literature), and we denote the critical exponents as
\begin{align}
\label{eq:lambda}
\lambda &=\begin{cases} \frac{2\tau-3}{\tau-2}  & \text{ if } 2< \tau < 4 \\
\frac52  &\text{ if } 4< \tau
 \end{cases}\\
 \label{eq:sigma}
\sigma & = \begin{cases} \frac{3-\tau}{\tau-2}  & \text{ if } 2< \tau < 3 \\
\frac{\tau-3}{\tau-2}  & \text{ if } 3< \tau < 4\\
\frac12   &\text{ if } 4< \tau\\
 \end{cases}\\
 \label{eq:beta}
\beta & = \begin{cases} \frac{1}{3-\tau} & \text{ if } 2< \tau < 3 \\
\frac{1}{\tau-3}  & \text{ if } 3< \tau < 4\\
1   &\text{ if } 4< \tau\\
 \end{cases}\\
 \label{eq:pc}
 p_c &=\begin{cases} \frac{1}{\frac{\tau-2}{3-\tau}d_{min}^{\tau-2}n^{\frac{3-\tau}{\tau-1}}-1} & \text{ if } 2<\tau<3\\
 \frac{1}{\frac{\tau-2}{\tau-3}d_{min}-1}  &\text{ if } \tau>3
 \end{cases}.
 \end{align}

\subsubsection{Proof of Theorem \ref{thm:weak_quan}}

In the proof, first we are going to sample the graph and percolate the edges, which gives a random graph with a component structure described in Section \ref{sec:prev_res}. Then, we are going to sample the seed sets, and we will understand which components the highest degree nodes and the uniform seed set are likely to ``hit'' (i.e., intersect).

The main difficulty of the proof is that in different parameter ranges, the highest degree nodes and the uniform seed set hit different types of clusters. We show that the highest degree nodes hit the components in decreasing order until a certain component size, which we call $C_{\mathrm{min}}^{(c)}$. For small $s$, $C_{\mathrm{min}}^{(c)}$ is exactly $C_1$, in which case the highest degree nodes are contained entirely in the giant, and for larger $s$, $C_{\mathrm{min}}^{(c)}$ is strictly smaller than $C_1$, in which case the highest degree nodes infect medium sized components in addition to the giant. We denote the contribution of these medium-sized components to the total size that the highest degree nodes infect by $E_1$ in the calculations below.

Similarly to the highest degree nodes, we denote the smallest component size for which all components of that size or larger are hit by the uniform seed set with high probability by $C_{\mathrm{min}}^{(u)}$. In contrast with the highest degree nodes, for small $s$, the uniformly selected seeds hitting the giant component becomes a rare event (occurring with probability $q_{p,1}=o(1)$). To rule out this rare event, which would skew the expected value, we condition on $\mathcal{E}$, the complement of this rare event (a standard technique in the theory of random graphs with heterogenous degree distribution). Thus, the $C_{\mathrm{min}}^{(u)}$ for small $s$ becomes undefined, and each uniformly selected seed hits a component with small expected size (denoted by $E_3$ below). As $s$ increases, the uniform seed set starts hitting the giant to give $q_{p,1}=\Theta(1)$ and $C_{\mathrm{min}}^{(u)}=C_1$. Increasing $s$ even further, similarly to the highest degree nodes, eventually we start having $C_{\mathrm{min}}^{(u)}<C_1$ and we denote by $E_2$ the contribution of these medium sized components to the total size of infected by the uniform seed set. 

Since $C_{\mathrm{min}}^{(c)}\le C_{\mathrm{min}}^{(u)}$, the only way the uniform seed set can infect more nodes than the highest $C_{\mathrm{min}}^{(c)} = C_{\mathrm{min}}^{(u)}=C_1$, because in this case all highest degree nodes are contained in the giant ($E_1=0$), but the uniform seed set can still hit some small components ($E_3>0$), which implies that the uniform seed set has a small advantage.

This intuition is made formal in the proof below. See Table \ref{tab:glossary} for a list of definitions used in the proof. In Claim \ref{lem:5eq}, in \eqref{eq:5eq1}-\eqref{eq:5eq5}, we explicitly derive rows 3-7 and 12-18 of Table \ref{tab:tau34}. Rows 8-9 and 19-20 of Table~\ref{tab:tau34} contain the statements of Theorem \ref{thm:weak_quan} and Corollary \ref{cor:weak_quan}. Entries of the rows 8-9 and 19-20 in Table \ref{tab:tau34} can be computed by substituting in entries from the previous rows into \eqref{eq:5eq6}, which we do in a case-by-case analysis after presenting the formal computations that support Claim \ref{lem:5eq}. 

\begin{notation}
In this section we drop the subscript $n$ from sequences $s$, $p$, $\theta$, and $\mathcal{E}$ and we use the simplified notation
\begin{align*}
    f_{n,\tau}(\theta,s) &= f_{\mathrm{CM}(n,\tau)}(p_{c,n,\tau}+\theta_n,s_n) \\
    f_{n,\tau}(\theta,s,\mathcal{E}) &=f_{\mathrm{CM}(n,\tau)}(p_{c,n,\tau}+\theta_n,s_n,\mathcal{E}_n).
\end{align*}
\end{notation}

\begin{claim}
\label{lem:5eq}
Under the assumptions $ 1 \gg \theta \gg n^{-(|\tau-3|)/(\tau-1)}$, $n \gg s \gg 1$ and definitions given in Table \ref{tab:glossary}, the following equations hold for $\tau\in(2,4)$
\begin{align}
\label{eq:5eq1}
q_{c,1} =\P(|\mathcal{CI}_0(s) \cap \mathcal{C}_1| >0)& = 1-O(n^{-\log(n)}),\\
\label{eq:5eq2}
q_{p,1}=\P(|\mathcal{UI}_0(s) \cap \mathcal{C}_1| >0) &= \begin{cases}
\Theta(s\theta^\ptaumTh) & \text{if }  s \ll \theta^{-\ptaumTh},\\
1-O(n^{-\log(n)}) & \text{if }  s \gg \theta^{-\ptaumTh},
\end{cases}\\
\label{eq:5eq3}
E_1 =\E \left [ \sum_{i=1}^{n_{c,p}} |\mathcal{C}_i| \mathbbm{1}_{\{ C_1>|\mathcal C_i| > C_{\mathrm{min}}^{(c)}\}} \right]&= \begin{cases}
0 & \text{if }  s \ll n \theta^{\frac{\tau-1}{|\tau-3|}},\\
\Theta(n^{1-\ptaumOn} s^{\ptaumOn}) & \text{if }  s \gg n \theta^{\frac{\tau-1}{|\tau-3|}},
\end{cases}\\
\label{eq:5eq4}
E_2 =\E \left [ \sum_{i=1}^{n_{c,p}} |\mathcal{C}_i| \mathbbm{1}_{\{ C_1>|\mathcal C_i| > C_{\mathrm{min}}^{(u)} \}}\right]&= \begin{cases}
0 & \text{if }  s \ll n \theta^{\frac{\tau-2}{|\tau-3|}},\\
\Theta(n^{\frac{\tau - 3}{\tau-2}}s^{\ptaumTw})  & \text{if }  s \gg n \theta^{\frac{\tau-2}{|\tau-3|}}.
\end{cases}\\
\label{eq:5eq5}
\end{align}
When $\tau\in(2,3)$, then
\begin{equation}
    E_3=\E_{u \sim \mathcal{U}(V)} \left [|\mathcal{C}(u)| \mathbbm{1}_{\{ |\mathcal C(u)|<  \mathrm{min}(C_{\mathrm{min}}^{(u)}, C_1)\}} \right]=\Theta(1),
\end{equation}
while for $\tau\in(3,4)$,
\begin{align}
    E_3 =\E_{u \sim \mathcal{U}(V)} \left [|\mathcal{C}(u)| \mathbbm{1}_{\{ |\mathcal C(u)|<  \mathrm{min}(C_{\mathrm{min}}^{(u)}, C_1)\}} \right]&= \begin{cases}
\Theta(\theta^{-1}) & \text{if }  s \ll n \theta^{\frac{\tau-2}{\tau - 3}},\\
\Theta\left(\left(\frac{n}{s}\right)^{\frac{\tau - 3}{\tau-2}} \right) & \text{if }  s \gg n \theta^{\frac{\tau-2}{\tau - 3}}.
\end{cases} 
\end{align}
Finally, for all cases, it holds that
\begin{equation}
\label{eq:5eq6}
f_{n,\tau}(\theta,s)=\frac{\E_{p_c+\theta} \big[ \mathbf {Cl}(\mathcal{CI}_0(s)) \big]}{\E_{p_c+\theta} \big[\mathbf{Cl} (\mathcal{UI}_0(s))  \big]}=\frac{ q_{c,1} C_1 + \Theta(E_1) + o(1/n) }{q_{p,1}C_1+ \Theta(E_2) + \Theta(sE_3)  }.
\end{equation}

\end{claim}
In what follows, we derive each equation in this claim. 

\textbf{Proof of \eqref{eq:5eq1}:}
Using Table \ref{tab:SF2}, we estimate the probability from below by the probability that the largest degree vertex $v_1$ with degree $\Theta(n^{1/(\tau-1)})$ is not in the giant. Here, we use the fact that the number of edges in the giant component is $\Theta(n)$. Therefore, in an exploration process of the giant component, we need to match the $\Theta(C_1)$ many half-edges, and none of these half-edges can be matched to $v_1$, which means that the probability that $v_1$ avoids the giant is $(1-d_1/n)^{\Theta(C_1)}$ with $C_1=\Theta(\theta^{1/(3-\tau)})$. This yields that
\begin{equation}
q_{c,1} > 1-  \P(v_1 \not \in \mathcal C_1) \approx 1-\left(1-\frac{d_1}{n} \right)^{\Theta(C_1)} \approx 1-\Theta\left(e^{-n^\ptaumOn\theta^\ptaumTh} \right) = 1-O(n^{-\log(n)})
\end{equation}
because we assumed $\theta \gg n^{-(|\tau-3|)/(\tau-1)}$.

\textbf{Proof of \eqref{eq:5eq2}:} 
Using that $C_1=\Theta(\theta^{1/(3-\tau)})$ from Table \ref{tab:SF2}, the probability that none of the uniformly selected seeds fall among the $C_1$ many vertices is
\begin{equation}
q_{p,1} = 1-\left(1-\frac{C_1}{n} \right)^s \approx 1-(1-\theta^\ptaumTh)^s  \approx 1-e^{-s\theta^\ptaumTh} \approx \begin{cases}
\Theta(s\theta^\ptaumTh) & \text{if }  s \ll \theta^{-\ptaumTh},\\
1-O(n^{-\log(n)}) & \text{if }  s \gg \theta^{-\ptaumTh}.
\end{cases}
\end{equation}

\textbf{Proof of \eqref{eq:5eq3}:}
We start by counting the number of half-edges incident to $\mathcal{CI}_0(s)=\{v_1, \dots, v_s\}$ as 

\begin{equation}
 H(\mathcal{CI}_0(s)):=\sum\limits_{i=1}^s d_i = \sum\limits_{i=1}^s \left(\frac{n}{i} \right)^\frac{1}{\tau-1} = n^\ptaumOn s^{1-\ptaumOn}.   
\end{equation}
Let us construct a (medium sized) component of given size $K$ using an exploration process, by matching half-edges one-by-one in the component. We must match $\Theta(K)$ half-edges, so the chance that none of these half-edges are mathced to the half-edges attached to vertices in $\mathcal{CI}_0(s)$  is
$$\P(\text{a half-edge is not matched with a half-edge attached to } \mathcal{CI}_0(s))=1- \frac{H(\mathcal{CI}_0(s))}{\Theta(n)} =1-\Theta\big((n/s)^{-(\tau-2)/(\tau-1)}\big),$$
where the denominator is $\Theta(n)$ since during the whole procedure the available total number of half-edges is $\Theta(n)$.
So, the probability that a component of size $K$ is not containing any of the vertices in $\mathcal{CI}_0(s)$ is 

\begin{equation}\label{eq:hitting-prob}
 \P(\mathcal{C}_u \cap \mathcal{CI}_0(s) = \emptyset| \mathcal C_u=K) = \exp\Big( -\Theta\big( K (\tfrac{n}{s})^{-(\tau-2)/(\tau-1)} 
\big)\Big).  
\end{equation}
Hence, components of size $K \gg (n/s)^\frac{\tau-2}{\tau-1} $ intersect with $\mathcal{CI}_0(s)$ with constant probability, whereas components of size $K \ll (n/s)^\frac{\tau-2}{\tau-1} $ do not. The threshold $(n/s)^\frac{\tau-2}{\tau-1}$ can either be larger than the size of the second largest component $C_2=\theta^{-(\tau-2)/(|\tau-3|)}$, in which case the entire $\mathcal{CI}_0(s)$ is contained in the giant component, or $(n/s)^\frac{\tau-2}{\tau-1}$ is smaller than $C_2$, in which case $\mathcal{CI}_0(s)$ hits some medium components as well. Solving $(n/s)^\frac{\tau-2}{\tau-1}< \theta^{-\frac{\tau-2}{|\tau-3|}}$ for the latter case, we get
\begin{equation}
\label{eq:ccmin}
C_{\mathrm{min}}^{(c)}=
\begin{cases} C_1 &\text{ if } s\ll n\theta^{\frac{\tau-1}{|\tau-3|}},\\
 \left(n/s \right)^\frac{\tau-2}{\tau-1} &\text{ if } s\gg n\theta^{\frac{\tau-1}{|\tau-3|}}. 
 \end{cases}
 \end{equation}
This implies that $s\ll n\theta^{\frac{\tau-1}{|\tau-3|}}$ we have $E_1=0$ with high probability. Otherwise, by \eqref{eq:hitting-prob}, we hit all components of size at least $C_{\min}^{(c)}$, and recalling that $n_{c,p}$ is the total number of percolated components, that is order $n$, we use 
by Table \ref{tab:SF2} for the distribution of component sizes to calculate  
\begin{align}
E_1&=\E \left [ \sum_{i=1}^{n_{c,p}} |\mathcal{C}_i| \mathbbm{1}_{\{ C_1>|\mathcal C_i| > C_{\mathrm{min}}^{(c)}\}} \right]
\approx n_{c,p} \sum\limits_{k=(n/s)^\frac{\tau-2}{\tau-1} }^{\theta^{-\frac{\tau-2}{|\tau-3|}}} k\cdot k^{-\frac{2|\tau-3|}{\tau-2}} \nonumber\\
&\approx n\int\limits_{(n/s)^\frac{\tau-2}{\tau-1} }^{\theta^{-\frac{\tau-2}{|\tau-3|}}} x^{1-\frac{2|\tau-3|}{\tau-2} } dx
\approx n \left(\frac{n}{s}\right)^{\frac{\tau-2}{\tau-1}\left(2-\frac{2|\tau-3|}{\tau-2} \right)  } 
= n^{\frac{\tau-2}{\tau-1}}s^{\ptaumOn},
\end{align}
because $1-\frac{2|\tau-3|}{\tau-2}=-\left(1+\ptaumTw\right)<-1$. We note that we used (and will use later) the simple result that $n_{c,p} = \Theta(n)$ because of the last row of Table \ref{tab:SF2} substituted with constant $k$.

\textbf{Proof of \eqref{eq:5eq4}:}
The expected number of uniformly chosen seeds in a cluster of size $K$ is $sK/n$, and similarly to \eqref{eq:hitting-prob}, the probability that $s$ uniformly chosen seeds avoid a cluster of size $K$ decays exponentially.  Hence, we expect the uniform seed set get all clusters with $K\gg \frac{n}{s}$ and some of the clusters with size $K\ll \frac{n}{s}$, which implies 
\begin{equation}
\label{eq:cpmin}
C_{\mathrm{min}}^{(u)}=
\begin{cases} \frac{n}{s} &\text{ if }  C_2 \gg \frac{n}{s},\\
C_1 &\text{ if } C_1 \gg \frac{n}{s} \gg C_2,\\
 \text{undefined} &\text{ if } \frac{n}{s} \gg C_1.
 \end{cases}
 \end{equation}
Then, if $s \ll n\theta^{\frac{\tau-2}{|\tau-3|}}$ we have  $C_{\mathrm{min}}^{(u)}$ equal $C_1$ or undefined, and therefore $E_2=0$. Otherwise, by Table \ref{tab:SF2},

\begin{equation}
E_2 = \E \left [ \sum_{i=1}^{n_{c,p}} |\mathcal{C}_i| \mathbbm{1}_{\{ C_1>|\mathcal C_i| > C_{\mathrm{min}}^{(u)} \}}\right]
\approx n_{c,p} \sum\limits_{k=n/s}^{\theta^{-\frac{\tau-2}{|\tau-3|}}} k\cdot k^{-\frac{2|\tau-3|}{\tau-2}} 
\approx n\int\limits_{n/s}^{\theta^{-\frac{\tau-2}{|\tau-3|}}}x^{1-\frac{2|\tau-3|}{\tau-2}} dx 
\approx n \left(\frac{n}{s}\right)^{2-\frac{2|\tau-3|}{\tau-2}} 
= n^{\frac{\tau - 3}{\tau-2}}s^{\ptaumTw}
\end{equation}
because $1-\frac{2|\tau-3|}{\tau-2}=-\left(1+\ptaumTw\right)<-1.$

\textbf{Proof of \eqref{eq:5eq5}:}
By Table \ref{tab:SF2} and \eqref{eq:cpmin},

\begin{equation}
\label{eq:E3compute}
E_3= \E_{u \sim \mathcal{U}(V)} \left [|\mathcal{C}(u)| \mathbbm{1}_{\{ |\mathcal{C}(u)|<  \mathrm{min}(C_{\mathrm{min}}^{(u)}, C_1)\}} \right]
\approx \sum\limits_{k=1}^{\mathrm{min}(C_{\mathrm{min}}^{(u)}, C_2)} k\cdot k^{-\frac{\tau-1}{\tau-2}} 
\approx \int\limits_{1}^{\mathrm{min}\left(\frac{n}{s},\theta^{-\frac{\tau-2}{|\tau-3|}} \right)} x^{-\ptaumTw}  dx.
\end{equation}
There are three cases for what the integral in \eqref{eq:E3compute} could evaluate to. If $\tau \in (2,3)$, then $-\ptaumTw<-1$ and $E_3=\Theta(1)$. In the other case, if $\tau \in (3,4)$, then $-\ptaumTw>-1$ and the integral in \eqref{eq:E3compute} evaluates to 
$$E_3=\mathrm{min}\left(\frac{n}{s},\theta^{-\frac{\tau-2}{|\tau-3|}} \right)^\frac{|\tau-3|}{\tau-2}.$$
Therefore, if $s \ll n\theta^{\frac{\tau-2}{|\tau-3|}}$ we have $E_3= \theta^{-1}$, otherwise, $E_3=\left(n/s\right)^{\frac{|\tau-3|}{\tau-2}}$

\textbf{Proof of \eqref{eq:5eq6}:}
We calculate the expected final size of the cluster of the uniform seed set first, i.e., the denominator in $f_{n,\tau}(\theta,s,\mathcal E_n)$. For the uniform seed set, following the definitions in Table \ref{tab:glossary}, since every cluster of size larger than $C_{\mathrm{min}}^{(u)}$ is hit by the uniform seed set with constant probability (hidden in the $\Theta$ notation before $E_2$),
\begin{equation}
    \E_p[\mathbf{Cl}(\mathcal{UI}_0(s))] = q_{p,1}C_1+ \Theta(E_2) + \E \left [ \sum\limits_{i=1}^{n_{c,p}} \mathcal C_i \mathbbm{1}_{\{\mathcal{UI}_0(s) \cap \mathcal{C}_i \ne \emptyset\}}\mathbbm{1}_{\{ \mathcal{C}_i < C_{\mathrm{min}}^{(u)} \}}  \right].
\end{equation}
Denote the last term on the right hand side by $T_3$. Since there are at most $s$ nodes in clusters that have size less than $C_{\mathrm{min}}^{(u)}$, and assuming each of these $s$ nodes hits a different cluster, we get the upper bound on the last term
$$T_3< sE_3.$$
For the lower bound on $T_3$, first we argue that $\Theta(s)$ seeds fall into these small components. Indeed, note that since $C_1+E_2 = o(n)$, we have
that the probability of a uniformly random chosen seed being in a cluster of size less than $C_{\mathrm{min}}^{(u)}$ is strictly positive (tends to one, in fact).
Moreover, since the expected number of uniformly chosen  seeds in a cluster of size $K \ll C_{\mathrm{min}}^{(u)}$ is $\frac{sK}{n} \rightarrow 0$, the probability of a uniformly random chosen seed being the only seed in its cluster also tends to 1. Thus, we can ignore the seeds colliding or falling in larger components, and we can write
 $$T_3 = \Omega(sE_3).$$
This establishes the bound on the denominator in \eqref{eq:5eq6}.

We continue with the numerator and estimate $\E_p[\mathbf{Cl}(\mathcal{CI}_0(s))]$, i.e., the cluster size of the highest degree nodes.  Following the definitions in Table \ref{tab:glossary}, we start with a lower bound that follows immediately from \eqref{eq:5eq1} and \eqref{eq:5eq3} and the fact that all components in $E_1$ will be infected with probability tending to  (derived in \eqref{eq:hitting-prob}):
$$\E_p[\mathbf{Cl}(\mathcal{CI}_0(s))] > q_{c,1} C_1 + \Theta( E_1).$$
It is left  to show an upper bound on $\E_p[\mathbf{Cl}(\mathcal{CI}_0(s))]$. Let us start with the case $s\ll n\theta^{\frac{\tau-1}{|\tau-3|}}$. In this case, using \eqref{eq:hitting-prob} and estimating each cluster-size trivially from above by $n$, we can write:  
\begin{align}
\label{eq:Cps_upper}
\E_p[\mathbf{Cl}(\mathcal{CI}_0(s))]| &< C_1 + \E \left[ \sum\limits_{i=1}^s \mathbbm{1}(v_i \not\in \mathcal C_1) \mathcal C(v) \right]  \nonumber \\
&< C_1 + \sum\limits_{i=1}^s \left( 1-\frac{C_1}{n} \right)^{d_i} n \nonumber \\
&< C_1 + \sum\limits_{i=1}^s \mathrm{exp} \left(-\left(\frac{n}{i}\right)^{\ptaumOn}\theta^\ptaumTh \right)n \nonumber \\
&< C_1+\mathrm{exp}\left(2\log(n)-\left(\frac{n}{s}\right)^{\ptaumOn}\theta^\ptaumTh \right).
\end{align}
Here we do a case distinction. Whenever   $s\ll n\theta^{\frac{\tau-1}{|\tau-3|}}$, we have
\begin{equation}\label{eq:small-error-condition}
2\log(n)-\left(\frac{n}{s}\right)^{\ptaumOn}\theta^\ptaumTh \rightarrow -\infty,    
\end{equation}
and thus $\E_p[\mathbf{Cl}(\mathcal{CI}_0(s))] < C_1 + o(1/n)$, which is what we needed, because in this case $E_1=0$ in \eqref{eq:5eq4}.

For the case $s\gg n\theta^{\frac{\tau-1}{|\tau-3|}}$ we use a coupling argument and monotonicity. Clearly, if we increase the edge-retention probability $p$ to $p'>p$, the total size of infected clusters cannot decrease.  So, let us consider percolation with $p'$ satisfying $\theta'=(s/n)^{\frac{|\tau-3|}{\tau-1}}\gg\theta$, implying that $p'>p$. Repeating \eqref{eq:Cps_upper} with $\theta'$, we get 
$$\E_p[\mathbf{Cl}(\mathcal{CI}_0(s))] < C_1' + o(1/n) = n\theta'^\ptaumTh  + o(1/n) = n^{1-\ptaumOn} s^{\ptaumOn}  + o(1/n),$$
since in this case \eqref{eq:small-error-condition} holds for the given choice of $s$ and $\theta'$.
Hence, by the monotonicity property of the cluster sizes in variable $p$ we arrive to
$$\E_p[\mathbf{Cl}(\mathcal{CI}_0(s))] < \E_{p'}[\mathbf{Cl}(\mathcal{CI}_0(s))] < n^{1-\ptaumOn} s^{\ptaumOn}  + o(1/n) = \Theta(E_1)+  o(1/n).$$

\begin{figure}[ht!]
  \begin{center}
  \includegraphics[width=\textwidth]{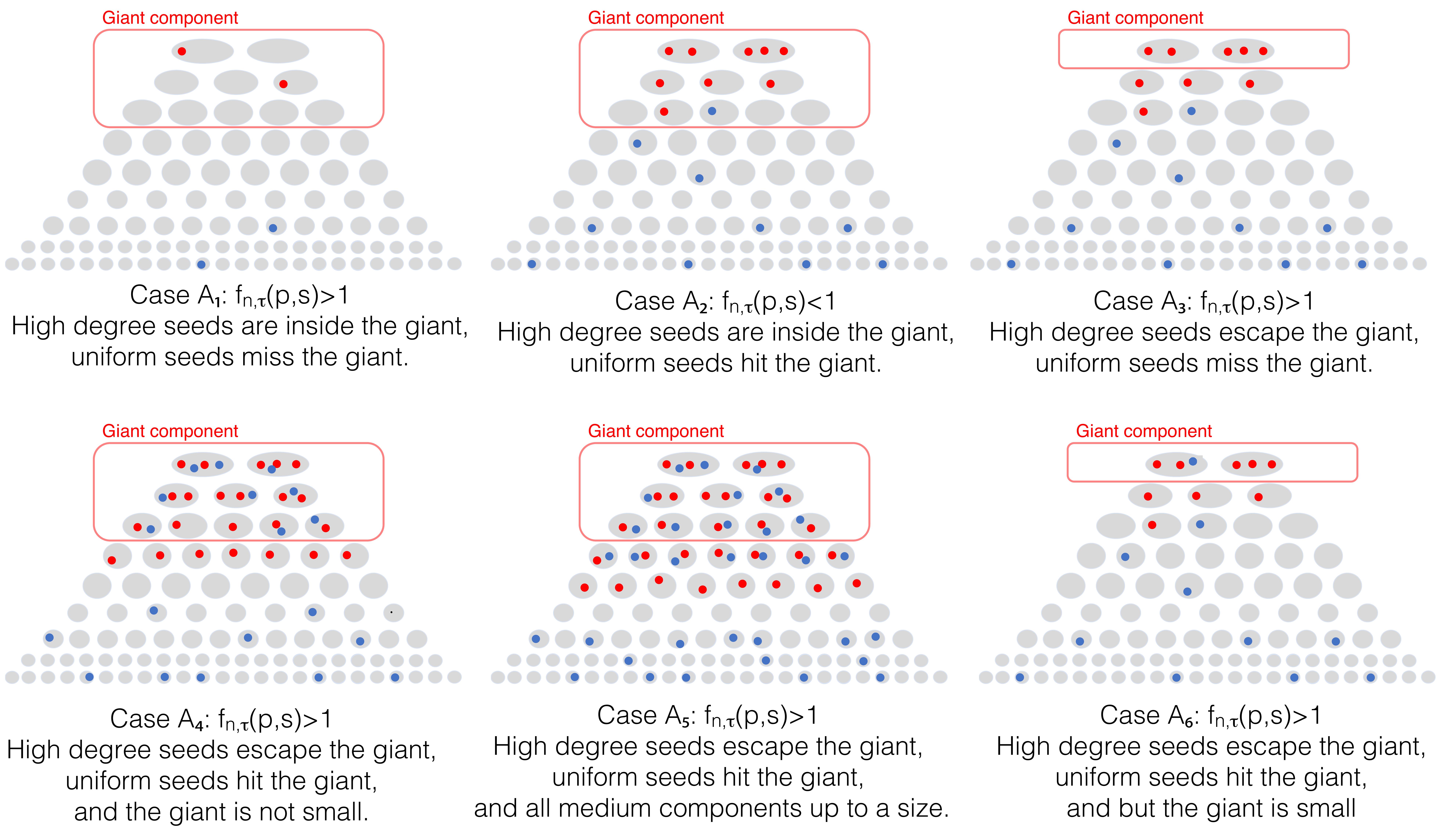}
  \end{center}
  \caption{\small Illustration for the derivation of Claim \ref{thm:weak_quan}. Each subfigure shows a schematic of each of the 6 parameter regions defined by $A_1$-$A_6$. The grey areas represent connected clusters in the percolated graph $G^p$, the red circles mark the $s$ highest degree nodes and the blue circles mark $s$ uniformly randomly chosen nodes.}
  \label{fig:cases5}
\end{figure}

\textbf{Completing the proof of Theorem \ref{thm:weak_quan} and Corollary \ref{cor:weak_quan}}:
We will use \eqref{eq:5eq6} or a conditioned version of it to compute $f_{n,\tau}(\theta,s,\mathcal{E})$. We treat each region $A_i$ in a case-by-case analysis to explain the final two rows of Table \ref{tab:tau34}. See Figure \ref{fig:cases5} for an illustration of each case.

\textbf{Case $A_1=\{ s \mid s \ll \mathrm{min}(\theta^{-\ptaumTh}, n \theta^{\frac{\tau-1}{|\tau-3|}})  \}$:}
Recall the high probability event $\mathcal E_n$ that $\mathcal{UI}_0(s)\cap C_1=\emptyset$ is required to hold on $A_1$. (Here, we assume $\mathcal E_n$ occurs and hence  work with a conditioned version of \eqref{eq:5eq6}). Since $\mathcal{E}$ only concerns the uniform seeds, and since $E_3$ and $E_2$ counting contributions of clusters avoiding the giant component $C_1$, the only term that needs to be changed in \eqref{eq:5eq6} is $q_{p,1}$, which needs to be changed to 0. For the other terms, by \eqref{eq:5eq1}-\eqref{eq:5eq5}, we have $q_{c,1}=1-O(n^{-\log(n)})$, $E_1=E_2=0$ and $E_3=\theta^{-\mathbbm{1}_{\tau \in (3,4)}}$,  so using \eqref{eq:5eq6} and the values from \eqref{eq:5eq1}--\eqref{eq:5eq5} 
\begin{equation}
f_{n,\tau}(\theta,s, \mathcal{E})=\frac{ q_{c,1} C_1 + o(1/n) }{\Theta(sE_3)  } =\Theta\left(\frac{n\theta^{\frac{1}{|\tau-3|}}}{s\theta^{-\mathbbm{1}_{\tau \in (3,4)}}}\right)=\Theta\left(\theta^{\frac{1}{|\tau-3|}+\mathbbm{1}_{\tau \in (3,4)}}ns^{-1} \right).
\end{equation}
In the normalized form, we get the linear relation $\tilde{f}_\tau(x,y,\mathcal{E})=1+\left(\frac{1}{|\tau-3|}+\mathbbm{1}_{\tau \in (3,4)} \right)x-y$. By \eqref{eq:5eq2}, 
$$\P(\mathcal{E})=1-q_{p,1}=1-\Theta(s\theta^{\ptaumTh}) \rightarrow 1$$
also holds.

\textbf{Case $A_2= \{ s \mid \theta^{-\ptaumTh}  \ll s \ll n \theta^{\frac{\tau-1}{|\tau-3|}}  \}$:}
By \eqref{eq:5eq1}-\eqref{eq:5eq5}, in this case, $q_{c,1}=1-O(n^{-\log(n)})$, $q_{p,1}=1-O(n^{-\log(n)})$, $E_1=E_2=0$ and $E_3=\theta^{-\mathbbm{1}_{\tau \in (3,4)}}$, which means that 
\begin{equation}
f_{n,\tau}(\theta,s)=\frac{(1-O(n^{-\log(n)}))C_1 +o(1/n)}{(1-O(n^{-\log(n)}))C_1+\Theta(s\theta^{-\mathbbm{1}_{\tau \in (3,4)}})}
=1-\Theta\left(\frac{s\theta^{-\mathbbm{1}_{\tau \in (3,4)}}}{C_1}\right)
=1-\Theta\left(\frac{\theta^{-\frac{1}{|\tau-3|}-\mathbbm{1}_{\tau \in (3,4)}}s}{n}\right).
\end{equation}
In the normalized form, we get  $\tilde{f}_\tau(x,y)=\left(\frac{1}{|\tau-3|}+\mathbbm{1}_{\tau \in (3,4)} \right) x-y$. The event $\mathcal{E}$ must occur in this case by definition, hence the results directly apply to $\tilde{f}_\tau(x,y,\mathcal{E})$ and $f_{n,\tau}(\theta,s,\mathcal{E})$ as well.

We note that this is the only case in which the uniform seed set infects more nodes than the highest degree nodes. As opposed to the other cases where we only had asymptotic results for $f_{n,\tau}(\theta,s)$, in this case we compute that the main order of the ratio is $1$, and even the asymptotics of the deviation from this main order. We can make such precise calculations only because both the numerator and the denominator of $f_{n,\tau}(\theta,s)$ are dominated by the expected size of the giant component, and these terms cancel each other. The deviation from $1$ then comes from the contribution of small clusters that the uniform seed set can infect. Intuitively, in this case a ``disassortative'' choice of seeds helps the infection to spread more.

\textbf{Case $A_3= \{ s \mid n \theta_n^{\frac{\tau-1}{|\tau-3|}} \ll s_n \ll  \theta_n^{-\ptaumTh}  \}$}: In this case, it is possible that event $\mathcal{E}$ does not occur (with some probability tending to $0$), and we have to work with a conditioned version of \eqref{eq:5eq6}. Since $\mathcal{E}$ only concerns the uniform seed set, and since $E_3$ and $E_2$ are conditioned on an event that implies $\mathcal{E}$, the only term that needs to be changed in \eqref{eq:5eq6} is $q_{p,1}$, which needs to be changed to 0. For the other terms, by \eqref{eq:5eq1}-\eqref{eq:5eq5}, we have $q_{c,1}=1-O(n^{-\log(n)})$, $E_1=n^{1-\ptaumOn} s^{\ptaumOn}$, $E_2=0$ and $E_3=\theta^{-\mathbbm{1}_{\tau \in (3,4)}}$, which means that 
\begin{equation}
f_{n,\tau}(\theta,s, \mathcal{E})=\Theta\left(\frac{n^{1-\ptaumOn} s^{\ptaumOn}}{s\theta^{-\mathbbm{1}_{\tau \in (3,4)}}}\right)=\Theta\left( \theta^{\mathbbm{1}_{\tau \in (3,4)}} \left(\frac{n}{s} \right)^{1-\ptaumOn} \right).
\end{equation}
In the normalized form, we get the linear relation $\tilde{f}_\tau(x,y,\mathcal{E})=\mathbbm{1}_{\tau \in (3,4)}x+\left(1-\ptaumOn\right)(1-y)$. By \eqref{eq:5eq2}, 
$$\P(\mathcal{E})=1-q_{p,1}=1-\Theta(s\theta^{\ptaumTh}) \rightarrow 1$$
also holds.

\textbf{Case
$A_4= \{ s \mid \mathrm{max}(\theta_n^{-\ptaumTh}, n \theta_n^{\frac{\tau-1}{|\tau-3|}}) \ll s_n \ll  
\mathrm{min}(n \theta_n^{\frac{\tau-2}{|\tau-3|}},
n \theta_n^{\frac{1}{|\tau-3|}})\}$:}
By \eqref{eq:5eq1}-\eqref{eq:5eq5}, in this case, $q_{c,1}=1-O(n^{-\log(n)})$, $q_{p,1}=1-O(n^{-\log(n)})$,  $E_1=n^{1-\ptaumOn} s^{\ptaumOn}$ $E_2=0$ and $E_3=\theta^{-\mathbbm{1}_{\tau \in (3,4)}}$, which means that
\begin{equation}
f_{n,\tau}(\theta,s)=\Theta\left(\frac{n^{1-\ptaumOn} s^{\ptaumOn}}{n\theta^{\frac{1}{|\tau-3|}}+s\theta^{-\mathbbm{1}_{\tau \in (3,4)}}}\right)
=\Theta\left( n^{-\ptaumOn}\theta^{-\ptaumTh}s^{\ptaumOn} \right)
\end{equation}
because $n\theta^{\frac{1}{|\tau-3|}}\gg s\theta^{-\mathbbm{1}_{\tau \in (3,4)}}$ holds due to $s \ll \mathrm{min}(n \theta_n^{\frac{\tau-2}{|\tau-3|}},
n \theta_n^{\frac{1}{|\tau-3|}})$ in the definition of $A_4$. In the normalized form, we get the linear relation $\tilde{f}_\tau(x,y)=-\ptaumTh x-\ptaumOn(1-y)$. The event $\mathcal{E}$ must occur in this case by definition, hence the results directly apply to $\tilde{f}_\tau(x,y,\mathcal{E})$ and $f_{n,\tau}(\theta,s,\mathcal{E})$ as well.

\textbf{Case $A_5= \{ s \mid n \theta^{\frac{\tau-2}{|\tau-3|}} \ll s  \}$}: By \eqref{eq:5eq1}-\eqref{eq:5eq5}, in this case, $q_{c,1}=1-O(n^{-\log(n)})$, $q_{p,1}=1-O(n^{-\log(n)})$,  $E_1=n^{1-\ptaumOn} s^{\ptaumOn}$, $E_2=n^\frac{\tau - 3}{\tau-2}s^{\frac{1}{\tau-2}}$ and $sE_3\le E_2$, which means that
\begin{equation}
f_{n,\tau}(\theta,s)=\Theta\left(\frac{n^{1-\ptaumOn} s^{\ptaumOn}}{n\theta^{\frac{1}{|\tau-3|}}+ n^{\frac{\tau - 3}{\tau-2}}s^{\ptaumTw}  }\right)
=\Theta\left( \left(\frac{n}{s}\right)^{\frac{1}{(\tau-1)(\tau-2)}} \right)
\end{equation}
because $n\theta^{\frac{1}{|\tau-3|}} \ll   n^{\frac{\tau-3}{\tau-2}}s^{\ptaumTw}$ holds due to $n \theta^{\frac{\tau-2}{|\tau-3|}} \ll s$ in the definition of $A_5$. In the normalized form, we get the linear relation $\tilde{f}_\tau(x,y)=\frac{1}{(\tau-1)(\tau-2)}(1-y)$. The event $\mathcal{E}$ must occur in this case by definition, hence the results directly apply to $\tilde{f}_\tau(x,y,\mathcal{E})$ and $f_{n,\tau}(\theta,s,\mathcal{E})$ as well.

\textbf{Case $A_6=\{ s \mid \mathrm{max}(\theta_n^{-\ptaumTh}, n \theta_n^{\frac{1}{|\tau-3|}}) \ll s_n \ll  
n \theta_n^{\frac{\tau-2}{|\tau-3|}}) \}:$ } In this case (which occurs only for $\tau \in (2,3)$), by \eqref{eq:5eq1}-\eqref{eq:5eq5}, we have $q_{c,1}=1-O(n^{-\log(n)})$, $q_{p,1}=1-O(n^{-\log(n)})$,  $E_1=n^{1-\ptaumOn} s^{\ptaumOn}$ $E_2=0$ and $E_3=\Theta(1)$, which means that
\begin{equation}
f_{n,\tau}(\theta,s)=\Theta\left(\frac{n^{1-\ptaumOn} s^{\ptaumOn}}{n\theta^{\frac{1}{|\tau-3|}}+s}\right)
=\Theta\left( \left(\frac{n}{s}\right)^{1-\ptaumOn} \right)
\end{equation}
because $n\theta^{\frac{1}{|\tau-3|}}\ll s$ holds due to  the definition of $A_6$. In the normalized form, we get the linear relation $\tilde{f}_\tau(x,y)=\left(1-\ptaumOn\right)(1-y)$. The event $\mathcal{E}$ must occur in this case by definition, hence the results directly apply to $\tilde{f}_\tau(x,y,\mathcal{E})$ and $f_{n,\tau}(\theta,s,\mathcal{E})$ as well.

\begin{table}
\centering
\footnotesize
\caption{\small Previous results for percolation with retention probability $p_n$ on a-geometric networks with power-law degree distributions (Configuration model $\mathrm{CM}(n,\tau)$) using definitions in \eqref{eq:lambda}, \eqref{eq:sigma}, \eqref{eq:beta}, and \eqref{eq:pc}. The second to last line is the size distribution of a uniformly random cluster, while the last line is the is the size distribution of the cluster of a uniformly random node.}
{\renewcommand{\arraystretch}{1.5}
\begin{tabular}{|l|c|c|c|}
\hline
parameter region & slightly subcritical  & critical window  & slightly supercritical  \\ \hline
$\mathrm{sign}(p_n-p_c)$ & $-$  & $-,0,+$  & $+$  \\ \hline
$\theta_n=|p_n-p_c|$ & $1 \gg \theta_n \gg n^{-\sigma/(\lambda-1)} $ & $n^{-\sigma/(\lambda-1)}\gg \theta_n$ &  $1 \gg \theta_n \gg n^{-\sigma/(\lambda-1)} $   \\ \hline \hline
$C_1$ &  $\theta_n^{-1/\sigma}$ & $n^{1/(\lambda-1)}$ &  $n\theta_n^\beta$  \\ \hline
$C_2$ & $\theta_n^{-1/\sigma}$  & $n^{1/(\lambda-1)}$  &  $\theta_n^{-1/\sigma}$   \\ \hline \hline
$\P_{i \sim \mathcal{U}(\{1, \dots, n_{c,p} \})}(\mathcal{C}_{i} =k \mid u\ne 1)$ & \multicolumn{3}{c|}{$ k^{-\lambda}e^{-k/\theta_n^{-1/\sigma}} $ }     \\ \hline
$\P_{u \sim \mathcal{U}(V)}(\mathcal{C}(u) =k \mid \mathcal{C}(u)\ne \mathcal{C}_1)$ &  \multicolumn{3}{c|}{$k^{-(\lambda-1)}e^{-k/\theta_n^{-1/\sigma}} $ }        \\ \hline
\end{tabular}}
\label{tab:SF1}
{\renewcommand{\arraystretch}{1.5}
\newline \vspace{10px}
\caption{Table \ref{tab:SF1} reparametrized with only the degree exponent $\tau$.}
\begin{tabular}{|l|c|c|c|}
\hline
parameter region & slightly subcritical  & critical window  & slightly supercritical  \\ \hline
$\mathrm{sign}(p_n-p_c)$ & $-$  & $-,0,+$  & $+$  \\ \hline
$\theta_n=|p_n-p_c|$ & $1 \gg \theta_n \gg n^{-\frac{|\tau-3|}{\tau-1}} $ & $n^{-\frac{|\tau-3|}{\tau-1}} \gg \theta_n$ &  $1 \gg \theta_n \gg n^{-\frac{|\tau-3|}{\tau-1}} $   \\ \hline \hline
$C_1$ &  $\theta_n^{-\frac{\tau-2}{|\tau-3|}}$ & $n^{\frac{\tau-2}{\tau-1}}$ &  $n\theta_n^{\frac{1}{|\tau-3|}}$  \\ \hline
$C_2$ & $\theta_n^{-\frac{\tau-2}{|\tau-3|}}$  & $n^{\frac{\tau-2}{\tau-1}}$  &  $\theta_n^{-\frac{\tau-2}{|\tau-3|}}$   \\ \hline \hline
$\P_{i \sim \mathcal{U}(\{1, \dots, n_{c,p} \})}(\mathcal{C}_{i} =k \mid u\ne 1)$ & \multicolumn{3}{c|}{$ k^{-\frac{2\tau-3}{\tau-2}}e^{-\frac{k}{C_1}}$ }     \\ \hline
$\P_{u \sim \mathcal{U}(V)}(\mathcal{C}(u) =k \mid \mathcal{C}(u)\ne \mathcal{C}_1)$ &  \multicolumn{3}{c|}{$k^{-\frac{\tau-1}{\tau-2}}e^{-\frac{k}{C_1}}$ }        \\ \hline
\end{tabular}}
\label{tab:SF2}
\end{table}

\begin{table}\centering
\footnotesize
\caption{\small Summary of the proof of Theorem \ref{thm:weak_quan}. See definitions for the notation in Table \ref{tab:glossary}. The colors of the columns $A_1$-$A_6$ are chosen to match Figure \ref{result_ranges}. Dark grey signifies the leading term of the numerator, light grey signifies the leading term of the denominator of $f_{n,\tau}(p,s)$. For all rows (except for $q_{c,1}$ and $q_{p,1}$, where the $O$ and $\Theta$ notation is made explicit) the values in the cells represent asymptotic values.}
\begin{tabular}{|l|c|c|c|c|c|c|}
\hline
$\tau$ & \multicolumn{6}{c|}{$\tau\in(3,4)$}    \\ \hline
$(s,\theta)$ & \yellow $A_1$ & \green$ A_2$ & \blue $A_3$  & \multicolumn{2}{c|}{\orange $A_4$} & \red $A_5$   \\ \hline \hline
$q_{c,1}$ & \multicolumn{6}{c|}{$1-O(n^{-\log(n)})$ }  \\ \hline
$C_{\mathrm{min}}^{(c)}$ &  \multicolumn{2}{c|}{\dgrey $C_1$ }  &   \multicolumn{4}{c|}{$(n/s)^\frac{\tau-2}{\tau-1}$}   \\ \hline
$E_1$ & \multicolumn{2}{c|}{$0$ }   &  \multicolumn{4}{c|}{ \dgrey $n^{1-\ptaumOn} s^{\ptaumOn}$ }   \\ \hline \hline
$q_{p,1}$ & $\Theta(s\theta^\ptaumTh)$ &  $1-O(n^{-\log(n)})$ &  $\Theta(s\theta^\ptaumTh)$ &  \multicolumn{3}{c|}{$1-O(n^{-\log(n)})$ }   \\ \hline
$C_{\mathrm{min}}^{(u)}$ &  $\emptyset$  & \lgrey $C_1$ & $\emptyset$  & \multicolumn{2}{c|}{\lgrey $C_1$ }   &   $n/s$    \\ \hline
$E_2$ & \multicolumn{5}{c|}{$0$ }  & $ \lgrey n^{\frac{\tau - 3}{\tau-2}}s^{\ptaumTw}$         \\ \hline
$sE_3$ & \lgrey $s\theta^{-1}$ & $s\theta^{-1}$ & \lgrey $s\theta^{-1}$ &\multicolumn{2}{c|}{ $s\theta^{-1}$ }    &   \lgrey  $s(n/s)^{\frac{\tau - 3}{\tau-2}}$     \\ \hline \hline
$f_{n,\tau}(\theta,s,\mathcal{E})$ &\yellow $\theta^{\frac{\tau-2}{\tau-3}}n/s$   & \green $1-\theta^{-\frac{\tau-2}{\tau-3}}s/n$ & \blue $\theta(n/s)^{1-\ptaumOn}$  & \multicolumn{2}{c|}{\orange $\theta^{-\ptaumTh}(n/s)^{-\ptaumOn}$} & \red  $(n/s)^{\frac{1}{(\tau-1)(\tau-2)}}$ \\ \hline
$\tilde{f}_\tau(\tilde{x},\tilde{y},\mathcal{E})$ &\yellow $1+{\frac{\tau-2}{\tau-3}}\tilde{x}-\tilde{y}$   & \green ${\frac{\tau-2}{\tau-3}}\tilde{x}-\tilde{y}$ &   \blue $\tilde{x}+\left(1-\ptaumOn\right)(1-\tilde{y})$ & \multicolumn{2}{c|}{ \orange $-\ptaumTh\tilde{x}-\ptaumOn(1-\tilde{y})$} & \red  $\frac{1}{(\tau-1)(\tau-2)}(1-\tilde{y})$ \\ \hline
\hline
$\tau$ & \multicolumn{6}{c|}{$\tau\in(2,3)$}    \\ \hline
$(s,\theta)$ & \yellow $A_1$ & \green$ A_2$ & \blue $A_3$  & \blue $A_6$  & \orange $A_4$ & \red $A_5$   \\ \hline \hline
$q_{c,1}$ & \multicolumn{6}{c|}{$\Omega(1)$ }  \\ \hline
$C_{\mathrm{min}}^{(c)}$ &  \multicolumn{2}{c|}{\dgrey$C_1$ }  &   \multicolumn{4}{c|}{$(n/s)^\frac{\tau-2}{\tau-1}$}   \\ \hline
$E_1$ & \multicolumn{2}{c|}{$0$ }   &  \multicolumn{4}{c|}{\dgrey $n^{1-\ptaumOn} s^{\ptaumOn}$ }   \\ \hline \hline
$q_{p,1}$ & $\Theta(s\theta^\ptaumTh)$ & $1-O(n^{-\log(n)})$ &  $\Theta(s\theta^\ptaumTh)$ &  \multicolumn{3}{c|}{$1-O(n^{-\log(n)})$ }   \\ \hline
$C_{\mathrm{min}}^{(u)}$ &  $\emptyset$  & \lgrey $C_1$ & $\emptyset$  & $C_1$ & \lgrey $C_1$   &  $n/s$    \\ \hline
$E_2$ & \multicolumn{5}{c|}{$0$ }  & \lgrey $n^{\frac{\tau-3}{\tau-2}}s^{\ptaumTw}$         \\ \hline
$sE_3$ & \lgrey $s$& $s$&  \multicolumn{2}{c|}{\lgrey $s$ }  &  \multicolumn{2}{c|}{$s$ }   \\ \hline \hline
$f_{n,\tau}(\theta,s,\mathcal{E})$ &\yellow $\theta^{\pThmtau}n/s$   & \green $1-\theta^{-\frac{1}{3-\tau}}s/n$ & 
\multicolumn{2}{c|}{\blue $(n/s)^{1-\ptaumOn}$} &
\orange $(n/s)^{-\ptaumOn}\theta^{-\pThmtau}$  &  \red $(n/s)^{\frac{1}{(\tau-1)(\tau-2)}}$  \\ \hline
$\tilde{f}_\tau(\tilde{x},\tilde{y},\mathcal{E})$ &\yellow $1+\pThmtau \tilde{x}-\tilde{y}$   & \green $\pThmtau\tilde{x}-\tilde{y}$ & \multicolumn{2}{c|}{\blue $\left(1-\ptaumOn\right)(1-\tilde{y})$}  &  \orange$-\pThmtau\tilde{x}-\ptaumOn(1-\tilde{y})$ & \red $\frac{1}{(\tau-1)(\tau-2)}(1-\tilde{y})$   \\ \hline
\end{tabular}
\label{tab:tau34}
\end{table}

\begin{table}\centering
\footnotesize
\caption{\small Definitions and glossary of notation}

\begin{tabular}{p{3.5cm}p{10cm}}
\hline
notation & definition/meaning   \\ \hline \hline
\multicolumn{2}{|c|}{ Network models }   \\ \hline \hline
$G=(V,E)$ & graph with node set $V$ and edge set $E$   \\ \hline
$[d_i]_{n}$ & expected degrees of the Configuration model \\ \hline
$v_i$ & node with the $i^{th}$ largest (expected) degree \\ \hline
$\tau$ & exponent of the power-law degree distribution \\ \hline
$\mathrm{CM}(n,\tau)$& Configuration model with size $n$ and degree exponent $\tau$ \\ \hline \hline

\multicolumn{2}{|c|}{ Bond percolation }   \\ \hline \hline

$p$ & bond percolation retention probability\\ \hline
$G^{p}$&  the bond percolated graph acquired by deleting each edge of $G$ with probability $1-p$\\ \hline
$p_c, p_{c,n,\tau}$ & critical point of bond percolation in general, and for the Configuration model $\mathrm{CM}(n,\tau)$\\ \hline
$\theta$ & $|p-p_c|$, deviation from critical point\\ \hline
$n_{c,p}$ & number of connected components of $G^{p}$\\ \hline
$\mathcal{C}_i$ & a set valued random variable that equals the $i^{th}$ largest component of $G^{p}$\\ \hline
$\mathcal{C}(u)$ & a set valued random variable that equals the component in $G^{p}$ which contains node $u$\\ \hline
$C_i$ & $\E[|\mathcal{C}_i|]$\\ \hline \hline

\multicolumn{2}{|c|}{ Seed selection strategies }   \\ \hline \hline
$s$ & the size of the seed set of an epidemic process \\ \hline
$\mathcal{CI}_0(s)$ & central area seed set of size $s$, the $s$ highest degree nodes \\ \hline
$\mathcal{UI}_0(s)$ & uniform seed set of size $s$ \\ \hline
$\mathbb E_p[\mathbf{Cl}(\mathcal{CI}_0(s))]$ & expected cluster size of the seed set with the $s$ highest degree nodes in the percolated graph with retention probability p\\ \hline
$\mathbb E_p[\mathbf{Cl}(\mathcal{UI}_0(s))]$ & expected cluster size of the seed set with the $s$ highest degree nodes in the percolated graph with retention probability p\\ \hline
$f_G(p,s)$ & $\frac{\mathbb E_p[\mathbf{Cl}(\mathcal{CI}_0(s))]}{\mathbb E_p[\mathbf{Cl}(\mathcal{UI}_0(s))]}$, expected final infection size ratio \\ \hline  \hline

\multicolumn{2}{|c|}{ Variables defined for the Configuration model}   \\ \hline \hline
$\lambda$, $\sigma$, $\beta$ & standard constants (depending on $\tau$) defined in \eqref{eq:lambda}, \eqref{eq:sigma} and \eqref{eq:beta}\\ \hline
$f_{n,\tau}(\theta,s)$ &$f_{\mathrm{CM}(n,\tau)}(p,s)$, expected final infection size ratio in the Configuration model \\ \hline
$\mathrm{dNorm}_n$ & $\log_n(x) \text{ if } x>1, -\log_n(1-x)-1 \text{ if } x<1$ \\ \hline
$\tilde{x},\tilde{y}, \tilde{f}_\tau(\tilde{x},\tilde{y}) $ & $\lim\limits_{n \rightarrow \infty} \log_n(\theta_n), \lim\limits_{n \rightarrow \infty} \log_n(s_n), \lim\limits_{n \rightarrow \infty} \mathrm{dNorm}_n(f_{n,\tau}(\theta,s))$ \\ \hline 
$\mathcal{E}$ & the event that either $s\in A_2 \cup A_4 \cup A_5 \cup A_6$ or  $s\in A_1 \cup A_3$ and $\mathcal{UI}_0(s) \cap \mathcal{C}_1 = \emptyset$ both hold \\ \hline
$\tilde{f}_\tau(\tilde{x},\tilde{y},\mathcal{E}), f_{n,\tau}(\theta,s,\mathcal{E})$ & the same as $\tilde{f}_\tau(\tilde{x},\tilde{y}), f_{n,\tau}(\theta,s)$ but conditioned on $\mathcal{E}$ \\ \hline  \hline
\multicolumn{2}{|c|}{ Variables defined in the analytic derivation }   \\ \hline \hline
$q_{c,1}$&$\P(|\mathcal{CI}_0(s) \cap \mathcal{C}_1| >0)$\\ \hline
$q_{p,1}$&$\P(|\mathcal{UI}_0(s) \cap \mathcal{C}_1| >0)$\\ \hline
$C_{\mathrm{min}}^{(c)}$ &$ \mathrm{min}\{ C_u \mid \P(|\mathcal{CI}_0(s) \cap \mathcal{C}_u| =\Theta(1)) \}$\\ \hline
$C_{\mathrm{min}}^{(u)}$ &$ \mathrm{min}\{ C_u \mid \P(|\mathcal{UI}_0(s) \cap \mathcal{C}_u| =\Theta(1)) \}$\\ \hline
$E_1$&$\E \left [ \sum_{i=1}^{n_{c,p}} |\mathcal{C}_i| \mathbbm{1}_{\{ C_1>|\mathcal C_i| > C_{\mathrm{min}}^{(c)}\}} \right]$\\ \hline
$E_2$&$\E \left [ \sum_{i=1}^{n_{c,p}} |\mathcal{C}_i| \mathbbm{1}_{\{ C_1>|\mathcal C_i| > C_{\mathrm{min}}^{(u)} \}}\right]$\\ \hline
$E_3$&$\E_{u \sim \mathcal{U}(V)} \left [|\mathcal{C}(u)| \mathbbm{1}_{\{ |\mathcal C(u)|<  \mathrm{min}(C_{\mathrm{min}}^{(u)}, C_1)\}} \right]$\\ \hline
$A_1$&$ \{ s \mid s \ll \mathrm{min}(\theta^{-\ptaumTh}, n \theta^{\frac{\tau-1}{|\tau-3|}})  \} $\\ \hline
$A_2$&$ \{ s \mid \theta^{-\ptaumTh}  \ll s \ll n \theta^{\frac{\tau-1}{|\tau-3|}}  \} $\\ \hline
$A_3$&$ \{ s \mid n \theta_n^{\frac{\tau-1}{|\tau-3|}} \ll s_n \ll  \theta_n^{-\ptaumTh}  \} \cup \{ s \mid \mathrm{max}(\theta_n^{-\ptaumTh}, n \theta_n^{\frac{\tau-2}{|\tau-3|}}) \ll s_n \ll  
n \theta_n^{\frac{1}{|\tau-3|}}) \}$\\ \hline
$A_4$&$ \{ s \mid \mathrm{max}(\theta_n^{-\ptaumTh}, n \theta_n^{\frac{\tau-1}{|\tau-3|}}) \ll s_n \ll  
\mathrm{min}(n \theta_n^{\frac{\tau-2}{|\tau-3|}},
n \theta_n^{\frac{1}{|\tau-3|}})\}$\\ \hline
$A_5$&$ \{ s \mid n \theta^{\frac{\tau-2}{|\tau-3|}} \ll s  \}$\\ \hline
$A_6$&$\{ s \mid \mathrm{max}(\theta_n^{-\ptaumTh}, n \theta_n^{\frac{1}{|\tau-3|}}) \ll s_n \ll  
n \theta_n^{\frac{\tau-2}{|\tau-3|}}) \}$\\ \hline
\end{tabular}
\label{tab:glossary}
\end{table}

\newpage


\end{document}